\documentclass[twocolumn]{aastex63}

\usepackage{float}
\usepackage{amsmath}
\usepackage{afterpage}
\usepackage{enumitem}
\usepackage{pifont}
\renewcommand\ion[2]{#1$\;${\scshape{#2}}}
\usepackage{statmath}

\def\co{$^{12}$CO}
\def\tco{$^{13}$CO}
\def\ceo{C$\,^{18}$O}

\def\htwo{H$_2$}
\def\water{H$_2$O}
\def\ammonia{NH$_3$}
\def\msun{M$_{\odot}$}
\def\lsun{L$_{\odot}$}

\def\kms{${\text{km s}^{-1}}$}

\def\hr{\ensuremath{^{\text{h}}}}
\def\min{\ensuremath{^{\text{m}}}}

\def\fsec{\ensuremath{\overset{{\text{s}}}{.}}}
\def\fdeg{\ensuremath{\overset{\circ}{.}}}

\def\farcs{\ensuremath{\overset{\prime\prime}{.}}}
\def\deg{\ensuremath{^{\circ}}}
\def\arcm{\ensuremath{^{\prime}}}
\def\arcs{\ensuremath{^{\prime\prime}}}

\def\aprox{$\sim$}
\def\mjy{mJy~beam$^{-1}$}
\def\Jyb{Jy~beam$^{-1}$}
\def\mic{$\mu$m}
\def\por{$\times$}
\def\isoss{J23053}

\defcitealias{Birkmann07}{Bir07}

\shorttitle{Collimated Flow from ISOSS J23053+5953 SMM2}
\shortauthors{Rodr\'iguez et al.}

\begin{document}

\title{Discovery of a Highly Collimated Flow from the High-Mass Protostar ISOSS J23053+5953 SMM2 }

\author{Tatiana M. Rodr\'iguez}
\affiliation{Physics Department, New Mexico Tech, 801 Leroy Pl., Socorro, NM 87801, USA.}

\author{Peter Hofner}
\affiliation{Physics Department, New Mexico Tech, 801 Leroy Pl., Socorro, NM 87801, USA.}
\affiliation{Adjunct Astronomer at the National Radio Astronomy Observatory, 1003 Lopezville Road, Socorro, NM 87801, USA.}

\author{Esteban D. Araya}
\affiliation{Physics Department, Western Illinois University, 1 University Circle, Macomb, IL 61455, USA.}

\author{Qizhou Zhang}
\affiliation{Center for Astrophysics—Harvard \& Smithsonian, 60 Garden Street, Cambridge, MA 02138, USA.}

\author{Hendrik Linz}
\affiliation{Max-Planck-Institut f\"{u}r Astronomie, K\"{o}nigstuhl 17, 69117 Heidelberg, Germany.}

\author{Stan Kurtz}
\affiliation{Instituto de Radioastronom\'ia y Astrof\'isica, Universidad Nacional Aut\'onoma de M\'exico, Apdo. Postal 72-3 (Xangari), Morelia, Michoac\'an 58089, Mexico.}

\author{Laura Gomez}
\affiliation{Joint ALMA Observatory, Alonso de C\'ordova 3107, Vitacura, Santiago, Chile.}

\author{Carlos Carrasco-Gonz\'alez}
\affiliation{Instituto de Radioastronom\'ia y Astrof\'isica, Universidad Nacional Aut\'onoma de M\'exico, Apdo. Postal 72-3 (Xangari), Morelia, Michoac\'an 58089, Mexico.}

\author{Viviana Rosero}
\affiliation{National Radio Astronomy Observatory, 1003 Lopezville Rd., Socorro, NM 87801, USA.}

\begin{abstract}

We present Very Large Array C, X, and Q-band continuum observations, as well as $1.3\,$mm continuum and CO(2-1) observations with the Submillimeter Array toward the high-mass protostellar candidate ISOSS J23053+5953 SMM2. Compact cm continuum emission was detected near the center of the SMM2 core with a spectral index of 0.24$\,(\pm0.15)$ between 6 and 3.6$\,$cm, and a radio luminosity of $1.3\,(\pm0.4)\,$mJy$\,$kpc$^2$. The $1.3\,$mm thermal dust emission indicates a mass of the SMM2 core of $45.8\,(\pm13.4)\,$M$_\odot$, and a density of $7.1\,(\pm1.2)\times 10^6\,$cm$^{-3}$. The CO(2-1) observations reveal a large, massive molecular outflow centered on the SMM2 core. This fast outflow ($>50\,$km $\,$s$^{-1}$ from the cloud systemic velocity) is highly collimated, with a broader, lower-velocity component. The large values for outflow mass ($45.2\,\pm12.6\,$M$_\odot$), and momentum rate ($6\,\pm2\times10^{-3}\,$M$_\odot\,$km$\,$s$^{-1}\,$yr$^{-1}$) derived from the CO emission are consistent with those of flows driven by high-mass YSOs. The dynamical timescale of the flow is between $1.5 - 7.2 \times 10^4\,$yr. We also found from the C$^{18}$O to thermal dust emission ratio that CO is depleted by a factor of about 20, possibly due to freeze out of CO molecules on dust grains. Our data are consistent with previous findings that ISOSS J23053+5953 SMM2 is an emerging high-mass protostar in an early phase of evolution, with an ionized jet, and a fast, highly collimated, and massive outflow.

\end{abstract}


\section{Introduction}
\label{sec:intro}

High-mass $(\text{M}\,>\,8\,\text{M}_{\odot})$ stars are of great importance for many astronomical research topics, yet despite recent significant progress, we still lack a detailed physical theory of their formation process (see e.g. the review by \citealt{Motte18}).
There are two major challenges in this research area: the great distance to massive star-forming regions, and the very short evolutionary time-scales.
The new generation of highly sensitive radio interferometers has mitigated  the former issue significantly, however, the latter is still difficult to surpass. 

One of the formation theories under debate is a scaled-up version of the process lower-mass objects undergo, namely, formation via disk accretion. Two phenomena are key ingredients in this scenario: jets of ionized material and molecular outflows, which contribute to the turbulence of the system and, most importantly, regulate the mass-loss process while shedding excess angular momentum and allowing accretion to proceed. Observationally, molecular outflows are ubiquitously detected toward high-mass young stellar objects (YSOs) \citep[e.g.,][]{Zhang01}, and there is no doubt that they play a key role in the formation process. In contrast, until recently, relatively few ionized jets were clearly identified toward high-mass YSOs \citep[e.g.,][]{Rosero16,Rosero19}.

To understand the role of the outflow/ionized jet phenomena in high-mass star formation, we need to study the youngest evolutionary phases, where the object is still in the process of accreting most of its final mass. While outflows have been associated with Ultracompact (UC) \ion{H}{ii} regions \citep[e.g.,][]{Shepherd96}, the flows are likely not actively driven anymore in the UC~\ion{H}{ii} phase. Prior to the formation of UC H~\textsc{ii} regions, the so called Hot Molecular Cores (HMCs) also drive molecular flows \citep[e.g.,][]{Araya2008,Hofner17}, but in this phase it is likely that nuclear burning has commenced and its energy output heats the surrounding matter to the observed high temperatures and large bolometric luminosities. In this paper we present the discovery of a flow/jet system in the source ISOSS J23053+5953 SMM2, which is a candidate for a high-mass YSO in an evolutionary phase clearly prior to the HMC phase.

ISOSS\footnote{ISO mission \citep{Kessler96}, ISOPHOT Serendipity Survey \citep[ISOSS,][]{Lemke96,bogun96}.} J23053+5953 (hereafter J23053), which coincides in position with IRAS 23032+5937, is a star-forming region located toward the Cepheus molecular complex at a distance of $4.3\pm 0.6\,$kpc \citep{Ragan12}. \cite{Wouterloot86} and \cite{Wouterloot88} detected \water\ and \ammonia\ masers (respectively) toward this region. A subsequent study by \cite{Wouterloot89} revealed the presence of a molecular outflow, thus confirming its star-forming nature. 
\cite{Bihr15} observed the \ammonia\ (1,1) and (2,2) lines towards \isoss\ and found a large scale abrupt radial velocity change with a
gradient larger than 30~\kms ~pc$^{-1}$, suggesting that cloud collision could have triggered the star formation in this region.

\cite{Birkmann07} (hereafter \citetalias{Birkmann07}) studied \isoss\ in a number of different molecular lines, as well as the mm/submm continuum\footnote{Note that \citetalias{Birkmann07} used a distance of $3.5\,$kpc, whereas we adopted a distance of $4.3\pm 0.6\,$kpc based on the more recent kinematic distance estimated by \cite{Ragan12}.}. They reported two submillimetric sources: SMM1 and SMM2.
In this work we will focus on SMM2, leaving the discussion of SMM1 to a future paper. From their thermal dust, and molecular line data
\citetalias{Birkmann07} estimated that the SMM2 core has a temperature of $17\,$K, a luminosity of $490\,$L$_\odot$, and a mass of 26~\msun , enclosed within \aprox8500~AU. Red-shifted HCO$^+$ absorption, and spectral energy distribution (SED) modeling indicated the presence of mass infall, with an accretion rate of $2.1\times 10^{-3}\, \text{M}_{\odot}\ \text{yr}^{-1}$, suggesting a collapsing core.
Their SED model further suggested that SMM2 is a massive cold molecular core with an emerging 4.6~\msun\ protostar, aged between 5000 and $3.6\times 10^4\,$yr, and they also showed that its luminosity is consistent with being entirely derived from accretion shocks. The single-dish data of \citetalias{Birkmann07} also indicated that an outflow is present in the region, which however was only marginally imaged by their interferometric data. Thus, although the current estimate for the mass and luminosity of the stellar object in SMM2 are characteristic for an intermediate-mass object, the \citetalias{Birkmann07} study strongly suggests that it is still assembling most of its future mass, and is therefore a prime candidate for a high-mass protostar.

Recently, \cite{Beuther21} reported high angular resolution and sensitivity observations with NOEMA toward J23053 to study the fragmentation during star formation in the region. They report 14 dust cores, where SMM2 corresponds to their core \#1, and at our angular resolution the weaker core \#5 also blends within our synthesized beam.

To continue the study of this promising system, we have obtained VLA observations in the cm continuum to search for an ionized jet, as well as SMA observations of the CO(2-1) transition to get more detailed information on the outflow in this region.
This paper is structured as follows: in Section~\ref{sec:observations} we present the VLA and SMA observational details, and describe the data reduction process. In Section~\ref{sec:results} we show the observational results, which we discuss in Section~\ref{sec:discussion}. Finally, in Section~\ref{sec:conclusions} we present a short summary of our most important results.

\section{Observations and Data Reduction}
\label{sec:observations}

\subsection{VLA Observations}

We observed the cm continuum emission from J23053 with NRAO's Karl G. Jansky
Very Large Array (VLA)\footnote{The National Radio Astronomy Observatory is a facility of the National Science Foundation 
operated under cooperative agreement by Associated Universities, Inc.} in the $6\,$cm (C), $3.6\,$cm (X), and $0.7\,$cm (Q) bands. 
Observational parameters are given in Table~\ref{tab:VLAobservations}. Note that for the C-band observations we combined data from 2 different observing dates.
For all observations we used a phase center of RA(J2000)~=~$23^h\, 05^m\, 23.0^s$, and Dec(J2000)~=~$59^\circ \,53^\prime \, 50\overset{\prime\prime}{.}0$, and the flux density calibration for all bands is based on observations of 3C48, using the "Perley-Butler 2010" flux scale \citep{PerleyButler2013}. 3C48 was also used for bandpass calibration, and the complex gains in C and X bands were derived from frequent observations of J2322+5057. For Q-band we used the source J2339+6010 to derive the complex gains, and this source was also used for pointing corrections in this band.

The data were reduced in the standard way using NRAO's Common Astronomy Software Applications (CASA, \citealt{casa}) package. Images with a variety of weighting schemes were made for each sub-band, 
as well as the combined bands at each wavelength. The resulting synthesized beams and rms noises for the naturally weighted combined maps are listed in columns 6 and 7
of Table~\ref{tab:VLAobservations}.

\begin{deluxetable*}{ccclccc}
\label{tab:VLAobservations}
\tabletypesize{\scriptsize}
\tablecaption{VLA Observational Parameters}
\tablecolumns{7}
\tablenum{1}
\tablewidth{0pt}
\tablehead{
\colhead{VLA Band} &
\colhead{$\nu_{central}$} &
\colhead{Bandwidth} & \colhead{Obs. Date} & \colhead{Configuration} & \colhead{Synth. Beam$^a$} & \colhead{Map rms$^a$}\\
\colhead{} & \colhead{[GHz]} &
\colhead{[GHz]} & \colhead{} &  \colhead{} & \colhead{[$^{\prime\prime}\,\, \times \,\, ^{\prime\prime}, \,\,^{\circ}$]} & \colhead{[$\mu$Jy beam$^{-1}$]}
}
\startdata
C$^b$ & 4.9, 7.4   & $2 \times 1$ & 2011 Aug 01 & A  & $0.43 \times 0.33, 65 $   &  1.6 \\
		&   &  &  2012 Oct 26  &   &    &   \\[3pt]
X		& 9.9, 11.4   & $2 \times 1$ & 2012 Oct 13 &    A  & $0.27 \times 0.19, 36 $   &  2.5 \\	
Q		& 41, 43,  45, 47 & $4 \times 2$ & 2013 Jun 08 &    C  & $0.66 \times 0.52, 26 $   &  17 \\		
	         \enddata
\tablenotetext{a}{For combined naturally weighted maps.}
\tablenotetext{b}{Combined data from 2 epochs.}
\end{deluxetable*}

\subsection{SMA Observations}
The observations were carried out on October 29, and November 9 2018 with the Submillimeter Array (SMA\footnote{The SMA is a joint project between the Smithsonian Astrophysical Observatory and the Academia Sinica Institute of Astronomy and Astrophysics, and is funded by the Smithsonian Institution and the Academia Sinica.})
in the compact configuration, which provides a maximum baseline of approximately 70 m. The phase center was set to 
RA(J2000)~=~$23^h\, 5^m\, 21.6^s$, and Dec(J2000)~=~$59^\circ \,53^\prime \, 43^{\prime\prime}$, and the total time on source was 13.5 hours. The SMA has eight 6-m antennas, and each antenna is equipped with four receivers, which allows dual frequency operation in pairs. We observed in the $1.3\,$mm band with central frequencies for the 230/240$\,$GHz receivers of 230.5 and 267.5$\,$GHz, respectively. 
At 1.3~mm, the SMA primary beam is \aprox 55\arcs, and the largest recoverable scale for the array in the selected configuration is about 20\arcs. 
We configured four 8 GHz wide sidebands with four spectral windows in each sideband. Initially, each 2 GHz spectral window had 16384 channels with a spectral resolution of $140\,$kHz, which subsequently was smoothed to 0.5 MHz.
Our bandpass contained a large number of molecular transitions, and we detected a total of 23 different spectral lines. In this paper we will concentrate on the emission from the three~CO isotopologues: $^{12}$CO(2-1), $^{13}$CO(2-1), and C$^{18}$O(2-1).

The planet Uranus was used for flux calibration, the quasar 3C84 was chosen as bandpass calibrator, and frequent observations of BL Lacertae and the quasar 0102+584 were used to calibrate the complex gains. 

After data calibration using the IDL-MIR package\footnote{The MIR cookbook by Charlie Qi can be found at \url{https://lweb.cfa.harvard.edu/~cqi/mircook.html}}, the data were transferred to CASA for imaging. We imaged the data of each day separately and, after careful inspection, both measurement sets were merged. Continuum subtraction was carried out using line free channels in each spectral window with the CASA task \texttt{uvcontsub}. We used all line free channels to form a continuum image for each side band, as well as a total continuum image using all 16 spectral windows, and a robustness value of 0.5.
%as well as a total 1.3 mm continuum image using all 16 spectral windows. 
The latter image corresponds to a central frequency of 240.7 GHz, with a total bandwidth of 3.7 GHz. The total 1.3 mm continuum image has a synthesized beam of 2\farcs7~$\times$~2\farcs4, at $\text{P.A.}=-$16\fdeg6, and a $\text{rms}=0.61\,$\mjy.

Spectral line cubes were made for each molecule using the task \texttt{tclean} and a robustness value of 0.5. The mean channel velocity width, synthesized beam size, and rms are 0.72~\kms, 3\farcs1~\por 2\farcs6, and 60~\mjy, respectively.

\section{Results}
\label{sec:results}

\subsection{VLA cm Data}
\label{subsec:VLA-results}

\begin{figure}
    \centering
    \includegraphics[width=\columnwidth]{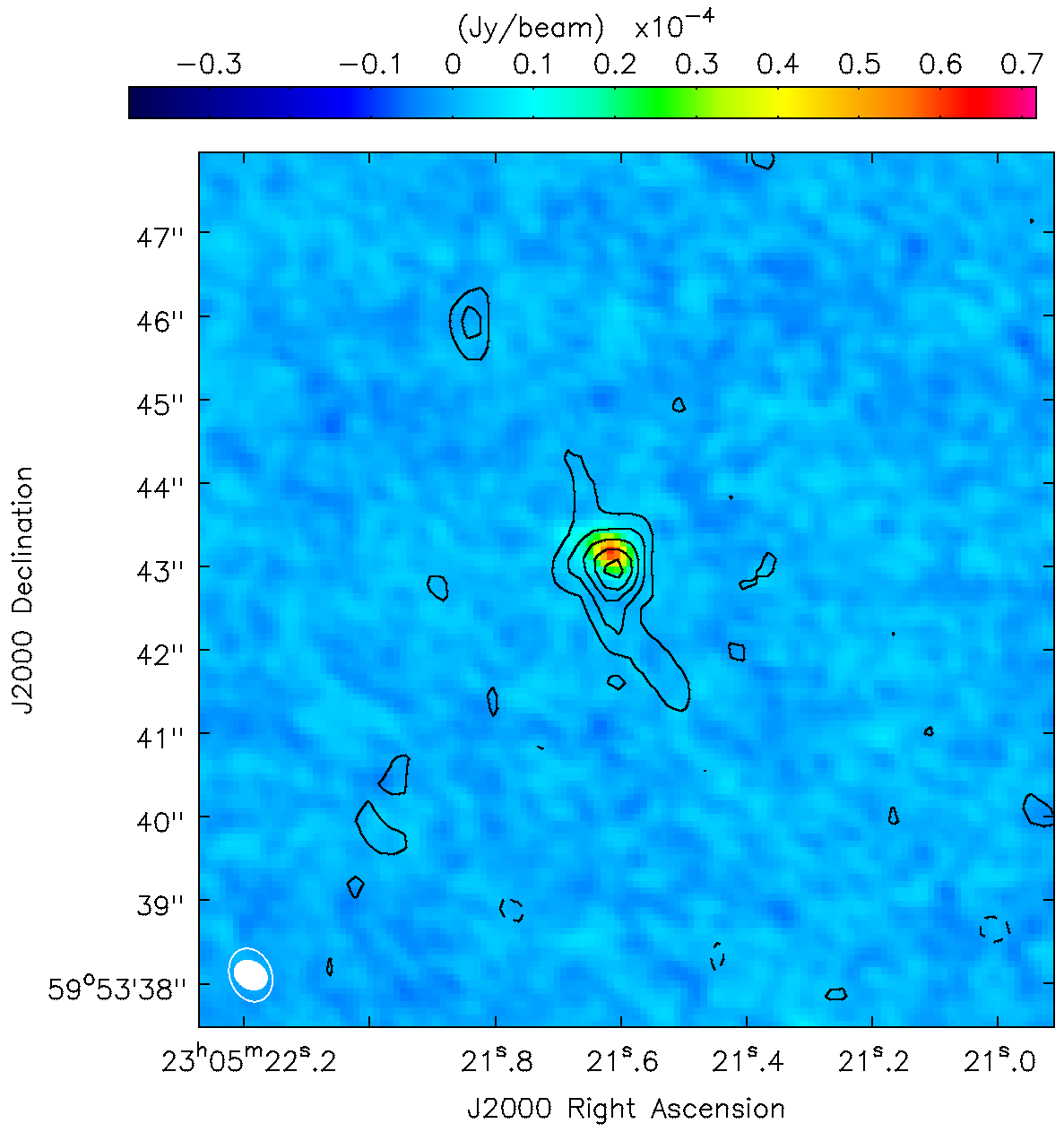}
    \caption{7~mm continuum emission (contours) overlaid on the 6~cm continuum
    emission (color) toward SMM2. The contours represent $-3, 3, 5, 7, 9 \text{ and } 11\sigma$ levels ($\sigma_{Q}=17\, \mu\text{Jy beam}^{-1}$), and the filled and contoured white ellipses in the bottom left show the 6~cm and 7~mm beam sizes, respectively.
    }
    \label{fig:CandQband}
\end{figure}

We detected the source SMM2 in all continuum bands. Figure~\ref{fig:CandQband} shows an
overlay of our C and Q-band images. At $6$ and $3.6\,$cm,
we detect a single compact source, but 2-D Gaussian fits indicate that
the emission is marginally resolved (see Table 2). The fitted peak position of the X-band emission is RA(J2000)~=~$23^h\, 05^m\, 21.61^s$, and Dec(J2000)~=~$59^\circ \,53^\prime \, 43\overset{\prime\prime}{.}2$. From the Gaussian fits we obtain the integrated fluxes in each sub-band, and we list these data in Table~2. 
This table also reports the deconvolved sizes and position angles (PAs).

From Figure~\ref{fig:CandQband} it is evident that the $7\,$mm emission is coincident with the cm source, and has a NE-SW extension.
We also note that the $7\,$mm peak is essentially coincident with the center of the $1.3\,$mm dust core (see below).

\begin{figure}
    \centering
    \includegraphics[width=\columnwidth]{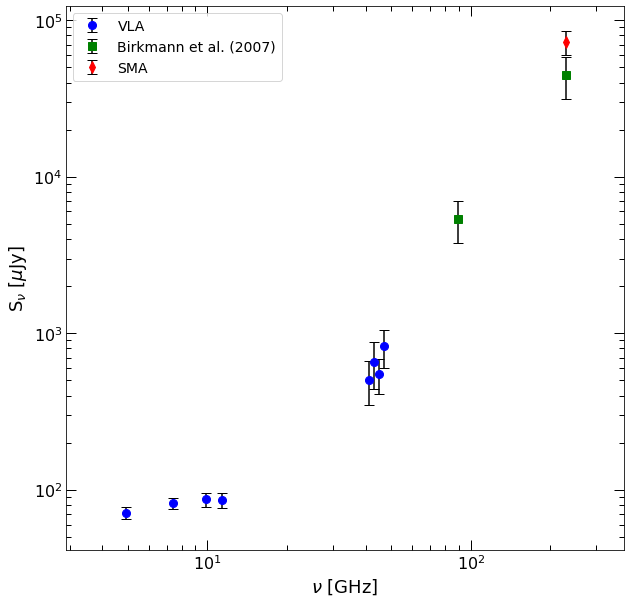}
    \caption{Spectral energy distribution of SMM2 using our VLA and SMA data, as well as the 3.4 and 1.3~mm data from \citetalias{Birkmann07}.}
    \label{fig:SED}
\end{figure}

In Figure~\ref{fig:SED} we show the cm/mm SED of SMM2, including our 1.3$\,$mm data, and also the 3.4 and 1.3$\,$mm data from \citetalias{Birkmann07}. 
The fluxes between C and X bands are rising, and a fit 
of the usual relation S$_\nu \propto \nu^\alpha$ results in a spectral index of
$\alpha = 0.24 \pm 0.15$. 
The fit was done through the Python SciPy \citep{Virtanen2020} task \texttt{curve\_fit}, which uses a linear least square algorithm.
This value is consistent with partially optically thick emission resulting from ionized gas with an electron density gradient, albeit smaller than the canonical value for ionized jets of $\alpha = 0.6$. 
At Q-band the flux begins to rise as expected from an extrapolation of the mm data, indicating that dust emission dominates at $7\,$mm.

\begin{deluxetable}{ccclccc}
\label{tab:VLA-gaussfit}
\tabletypesize{\scriptsize}
\tablecaption{SMM2 cm Source Properties}
\tablenum{2}
\tablecolumns{4}
\tablewidth{0pt}
\tablehead{
\colhead{$\nu_{c}$} &
\colhead{S$_\nu$}  & \colhead{$\theta_{maj} \times \theta_{min}$} & \colhead{PA}\\
\colhead{[GHz]} &
\colhead{[$\mu$Jy]}  & \colhead{[mas $\times$ mas]} & \colhead{[$^\circ$]}
}
\startdata
4.9	& 71.2 (5.2)   & 302 (75)  $\times$ 170 (95) & 57 (41)  \\
7.4	&  82.4 (5.1)  &  311 (35)  $\times$  193 (30) & 57 (13)  \\
9.9  &  86.9 (7.7)  &  212 (45)  $\times$  113 (40) &  23 (23)  \\
11.4 &  85.7 (8.5)  &  161 (45)  $\times$  76 (50) &  47 (29)  \\
 41.0     & 504 (149)  & 1100 (380)  $\times$  730 (300) & 26 (44)  \\
 43.0     & 659 (210)  & 1150 (460)  $\times$  850 (480) & 34 (61)  \\
 45.0     & 547 (128)  &  620 (230)  $\times$ 575 (325) & 102 (113)  \\
 47.0     &  828 (211)  & 1100 (350)  $\times$ 490 (240) & 1(24)  \\[3pt]
\enddata
\tablenotetext{}{Values in parentheses are formal errors from the 2-D Gaussian fits.}
\end{deluxetable}

%======================================================================-
%======================================================================-
\subsection {SMA 1.3 mm Continuum}

\begin{figure}
    \centering
    \includegraphics[width=\columnwidth]{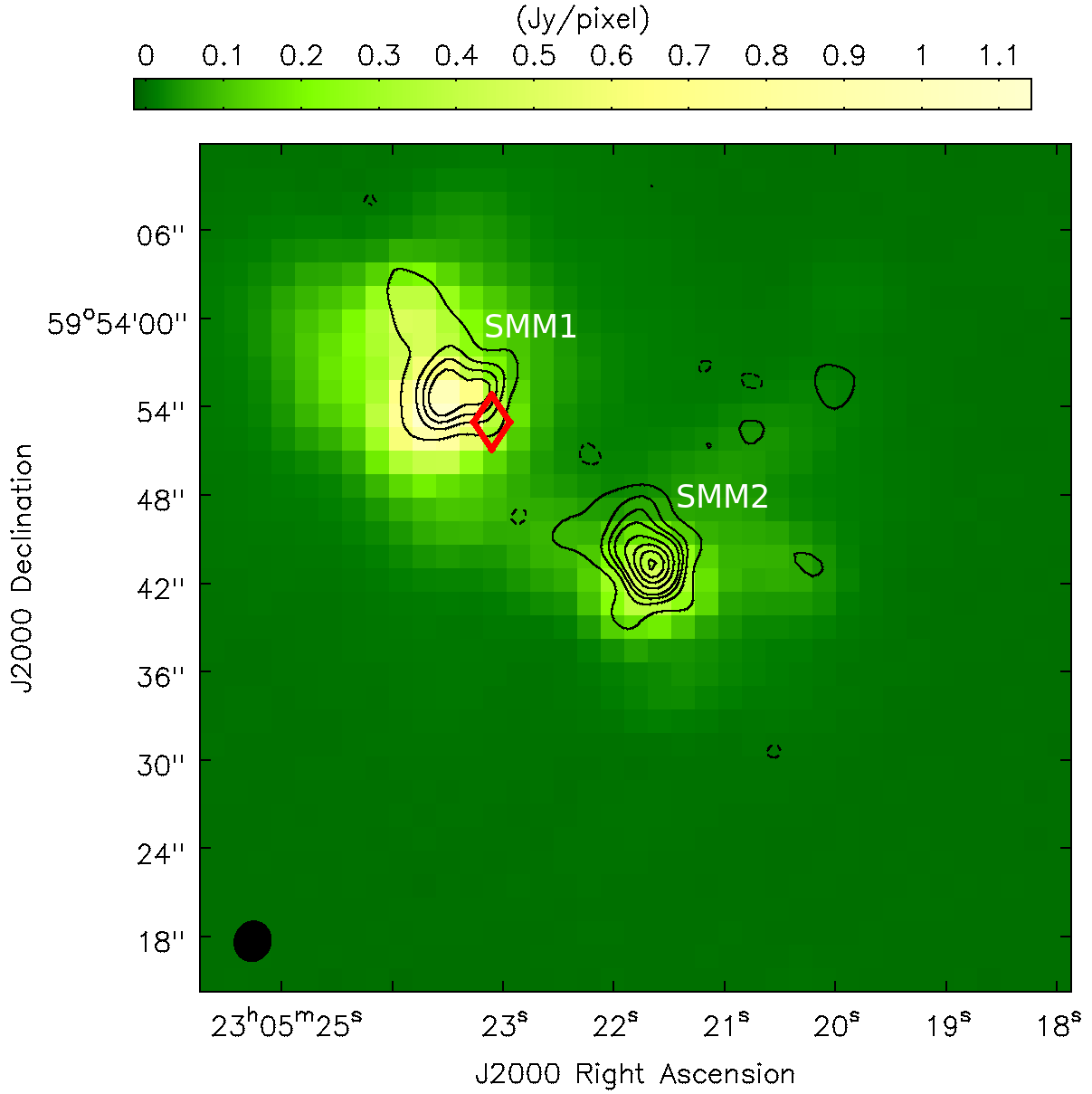}
    \caption{SMA 1.3 mm continuum emission (black contours) overlaid on Herschel 70 \mic\ emission (color scale) toward \isoss. The contours indicate -4, 4, 10, 15, 20, 30, 40, 50 and 60$\sigma$ levels ($\sigma$=0.61~\mjy). The northeast source is SMM1, and the source located to the center of the figure is SMM2. The red diamond shows the position of IRAS 23032+5937, and the black ellipse in the bottom left of the image shows the 1.3 mm beam size.}
    \label{fig:cont-70mic}
\end{figure}

\begin{figure}%[b]
    \centering
    \includegraphics[width=\columnwidth]{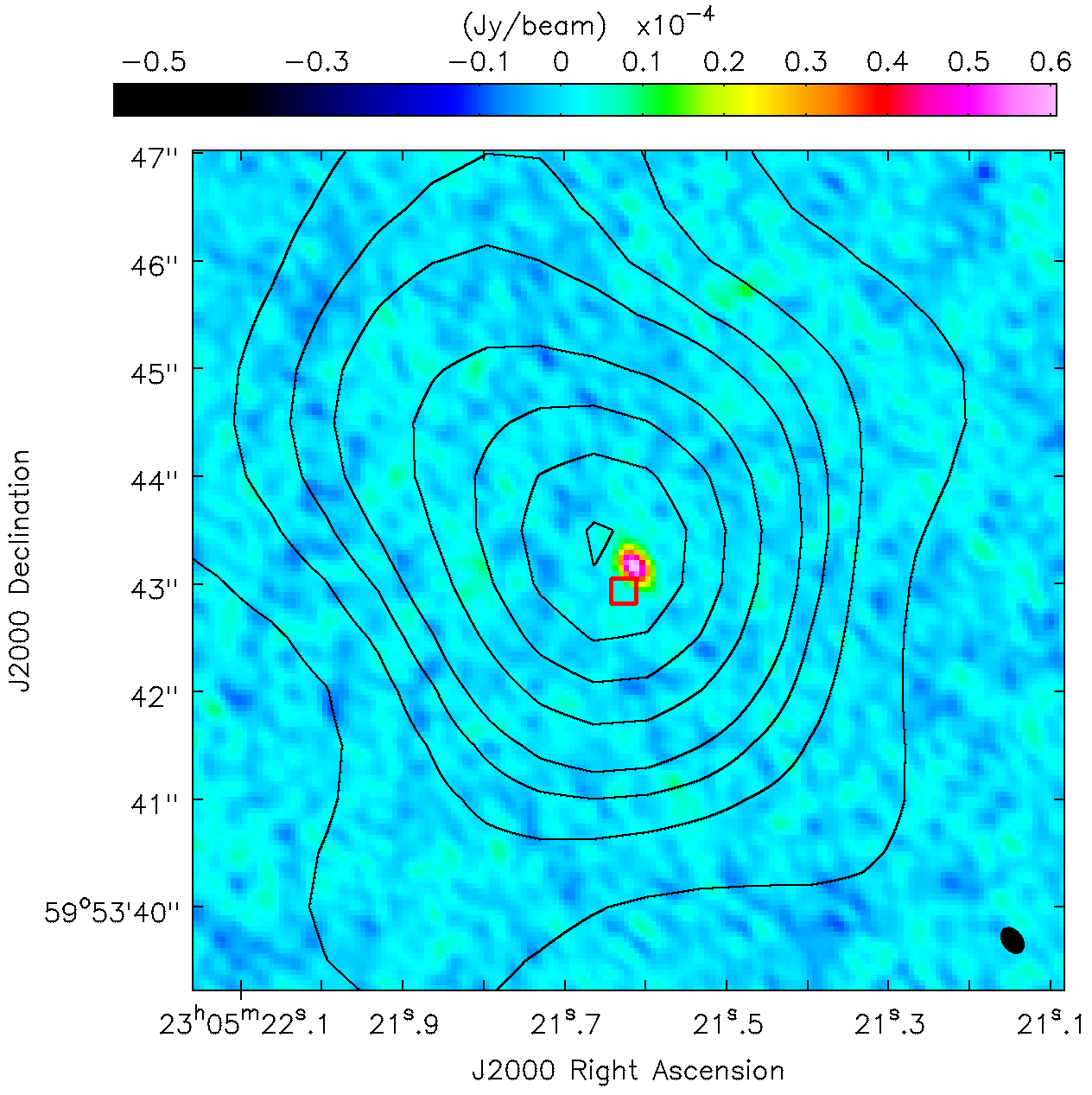}
    \caption{3.6 cm emission (color) toward SMM2. Black contours show the 1.3~mm emission, with the same contour levels as in Figure~\ref{fig:cont-70mic}. The red square indicates the 7~mm peak position, and the black ellipse at the bottom right represents the 3.6 cm beam size.}
    \label{fig:cont-Xband}
\end{figure}

In Figure \ref{fig:cont-70mic} we show the 1.3 mm continuum emission toward J23053 in black contours, overlaid on a Herschel 70~\mic\ image.
The northeastern source is SMM1, the source near the center of the figure is SMM2, and the red diamond indicates the nominal position of IRAS$\,$23032+5937. With our angular resolution, SMM2 has a compact but slightly resolved component, within a larger, elongated emission in the NE-SW direction.
In Figure~\ref{fig:cont-Xband} we show the 1.3~mm emission in contours overlaid on the 3.6~cm emission towards SMM2. We have also marked with a red square the position of the 7~mm peak.

We fitted a 2-D Gaussian model to the $1.3\,$mm continuum emission of SMM2 using the CASA task \texttt{imfit}, and obtained a deconvolved size of $3\overset{\prime\prime}{.}7 \,(\pm0.4) \times 1\overset{\prime\prime}{.}9 \,(\pm0.4)$ (which at 4.3~kpc translates to linear sizes of approximately $16000\times 8000\,$ au), a center position of RA(J2000)~=~23\hr05\min21\fsec6, Dec(J2000)~=~59\deg53\arcm43\arcs, and a PA of 32\fdeg8$\,(\pm9\overset{\circ}{.}7)$. The integrated $1.3\,$mm flux of SMM2 from the Gaussian fit is $71.5 \pm 6.7\,$mJy.
This is larger than the $45\,$mJy reported by \citetalias{Birkmann07}, who observed the region with a higher angular resolution (1\farcs15 \por\ 0\farcs87). The difference between these measurements is likely due to our observations detecting more extended emission with a larger synthesized beam.

We used once again the SciPy task \texttt{curve\_fit} and fitted the mm-SED composed of our $1.3\,$mm data combined with the flux densities reported by \citetalias{Birkmann07} at 1.3 and 3.4 mm emission (assuming an error of 30\% in their measured fluxes), as well as the VLA 7~mm fluxes. We obtained a spectral index of $\alpha=2.7\, \pm0.1$, which indicates optically thin dust emission.
Using the formalism of \cite{Hildebrand83} we thus calculate the total core mass with the expression
\begin{equation}
M_d=\frac{d^2S_{\nu}R_g}{B_{\nu}(T_d)\kappa_{\nu}},
\end{equation}
where \textit{d} is the distance to the source, $S_{\nu}$ is the integrated flux, $R_g$ is the gas-to-dust ratio, $B_{\nu}(T_d)$ is the Planck function, and $\kappa_{\nu}$ is the dust opacity. 
We used a gas-to-dust ratio of 150, higher than the usual value of 100, in order to account for the change of this ratio throughout the Galaxy, having in mind that SMM2 is located at a galactocentric radius $\gtrsim10\,$kpc \citep[see][]{Gianetti17, Bosco19}.
For the temperature, given that we are probably detecting a considerable amount of cold dust, we used the value estimated by \citetalias{Birkmann07} of $T_{cold}=17.3$ K. We used an opacity value for dust with thick ice mantles and densities on the order of $10^6\ \text{cm}^{-3}$ from the work of \citet[][Table 1 Column 6]{Ossenkopf94}, i.e. \ensuremath{\kappa_{1.3mm}=0.962\ \text{cm}^2\ \text{g}^{-1}}. With the parameters stated, we obtain a total mass for the clump of 45.8$\,\pm 13.4$~\msun.
Using this result, and modeling the dust core as a constant density sphere, we estimate a total density  of $n(H_2)=7.1\, (\pm 1.2)\times 10^6\ \text{cm}^{-3}$, and a column density of
$N(H_2) =1.2\,(\pm 0.1)\times 10^{24}\ \text{cm}^{-2}$. In Table \ref{tab:cont-par} we summarize the core parameters derived from the 1.3 mm continuum emission.

\begin{deluxetable*}{c c c c c c}
\tablenum{3}
\tablecaption{SMM2 1.3~mm Derived Parameters\label{tab:cont-par}}
\tabletypesize{\scriptsize}
\tablewidth{0pt}
\tablehead{
\colhead{Deconvolved size} & \colhead{S$_{\nu}$} & \colhead{T$_d$} & \colhead{M} & \colhead{n(H$_2$)} & \colhead{N(H$_2$)}\\
\colhead{[$\,$\arcs\ \por\ \arcs , \deg$\,$]} & \colhead{[mJy]} & \colhead{[K]} & \colhead{[\msun]} & \colhead{[10$^{6}\ \text{cm}^{-3}$]}& \colhead{[10$^{24}\ \text{cm}^{-2}$]}
}
\startdata
3.68$\,\pm$0.45 \por\ 1.9$\,\pm$0.42, 32.8$\,\pm$9.7 & 71.5$\,\pm$6.7 & 17.3 & 45.8$\,\pm$13.4 & 7.1$\,\pm$1.2 & 1.2$\,\pm$0.1
\enddata
\end{deluxetable*}

%======================================================================-
%======================================================================-
\subsection{CO(2-1), $^{13}$CO(2-1) and C$\,^{18}$O(2-1)}
\label{subsec:results-co}

\begin{figure}
    \centering
    \includegraphics[width=0.95\columnwidth]{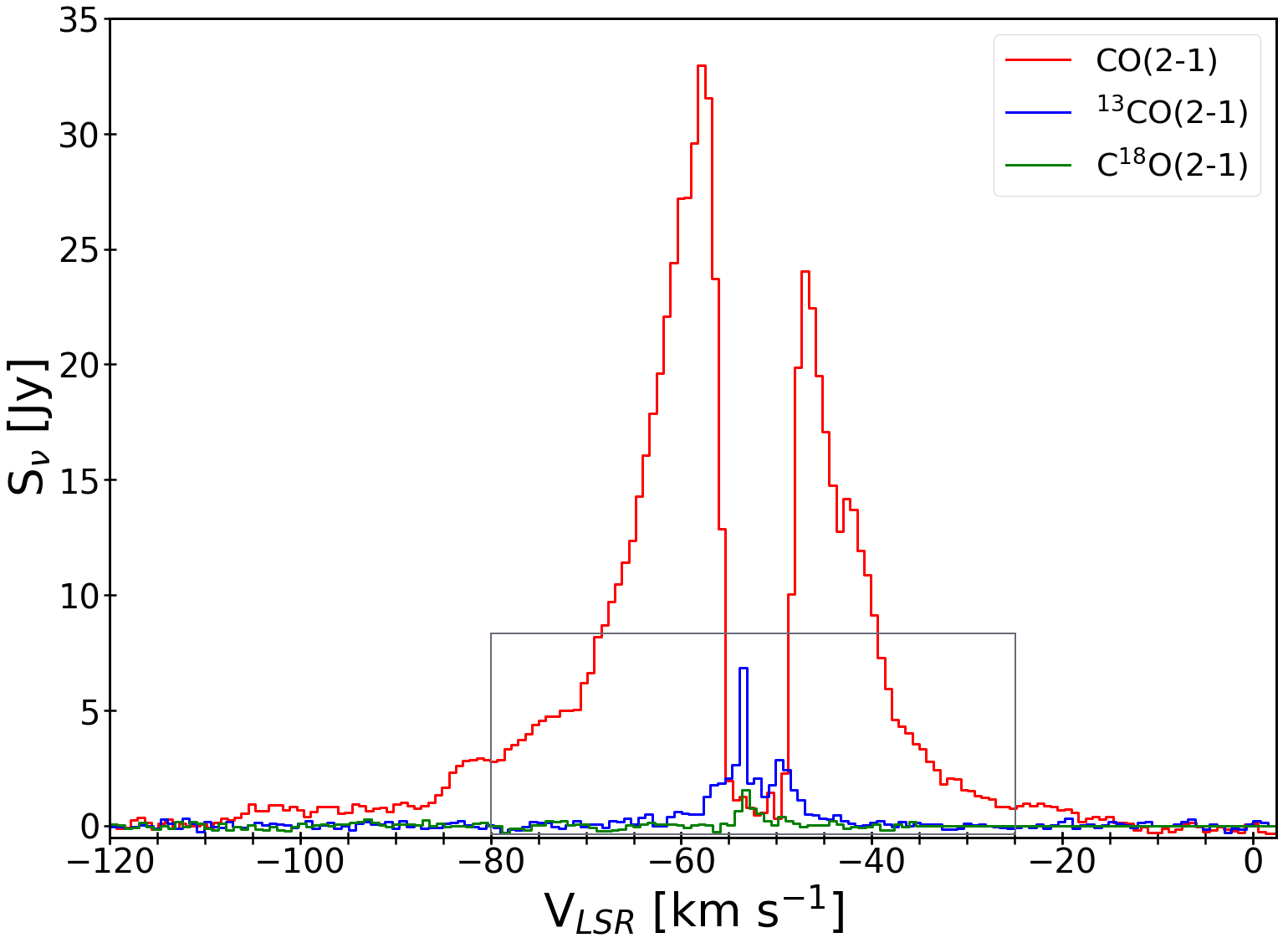}
    \includegraphics[width=0.95\columnwidth]{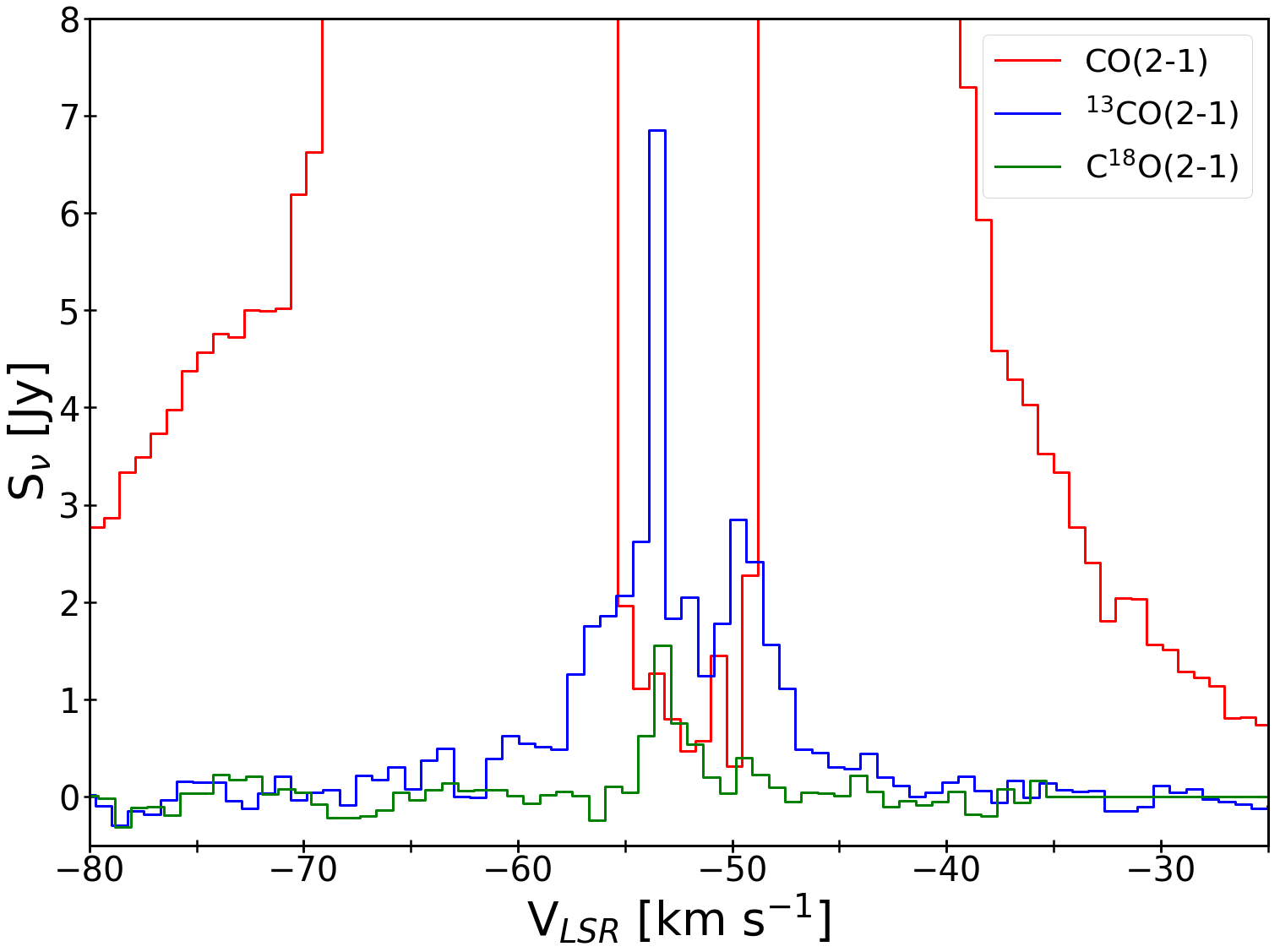}
    \caption{\textit{Top}: Integrated $J=2$-1 spectra of three CO isotopologues toward SMM2. The CO(2-1), \tco(2-1) and \ceo(2-1) spectral lines are shown in red, blue and green, respectively. The gray rectangle marks the region shown in the bottom panel. \textit{Bottom}: Zoomed-in version of the top panel, with a focus on the \tco\ and \ceo\ spectra.}
    \label{fig:CO-spectra}
\end{figure}

In Figure \ref{fig:CO-spectra} we show the integrated spectra of the CO(2-1), \tco(2-1), and \ceo(2-1) line emission associated with SMM2.
The CO line shows prodigious wings that extend from $-$108.3 to $-$13.2 \kms, and a deep, central self-absorption feature. A similar structure, albeit to a lesser extent, is observed in \tco (2-1).
For this line, the wing emission extends from $-61.5$ to $-47$~\kms. In contrast, there is no evidence for line wings in the \ceo\ spectrum. A Gaussian fit to the \ceo(2-1) line using the Astropy\footnote{\url{http://www.astropy.org} a community-developed core Python package for Astronomy \citep{astropy13, astropy18}.} subpackage \texttt{modeling} results in a mean velocity of $-52.7\,\pm0.3$~\kms. We will take this number as the systemic velocity for the SMM2 core.

\begin{figure*}
     \centering
     \includegraphics[width=0.96\textwidth]{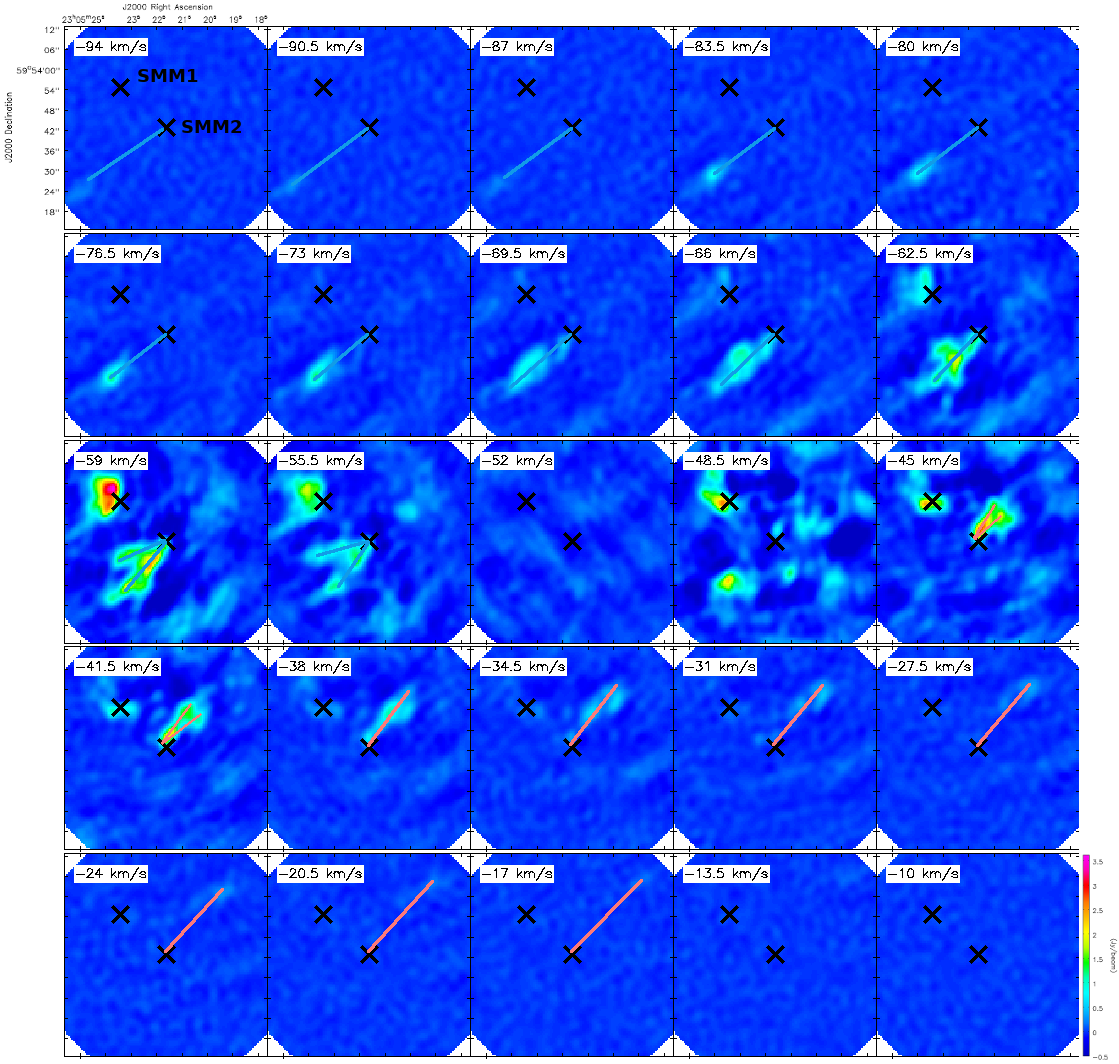}
     \caption{Channel maps of the CO(2-1) emission toward \isoss. The emission in each panel was summed over 5 channels ($\Delta v=3.5\,$\kms), resulting in a mean channel rms of $\sim$0.3 \Jyb\ for the channels with velocity between $-62.5$ and $-41.5$ \kms, and      $\sim$30 \mjy\ for the rest. The black crosses to the NE and to the center of each panel mark the positions of the SMM1 and SMM2 dust cores, respectively. The blue and red lines were drawn to indicate the flow direction. We can see bipolar emission arising from SMM2 in the SE-NW direction.}
     \label{fig:CO-vel-channel}
 \end{figure*}

In Figure~\ref{fig:CO-vel-channel} we show channel maps of the $^{12}$CO(2-1) transition.
The emission is clearly
centered on SMM2 and has a narrow bipolar morphology oriented in the SE-NW direction with a position angle of about 130\deg, which is approximately perpendicular to the $1.3$~mm continuum emission shown in Fig.~\ref{fig:cont-70mic}.

\begin{figure*}%[!ht]
     \centering
     \includegraphics[width=0.98\textwidth]{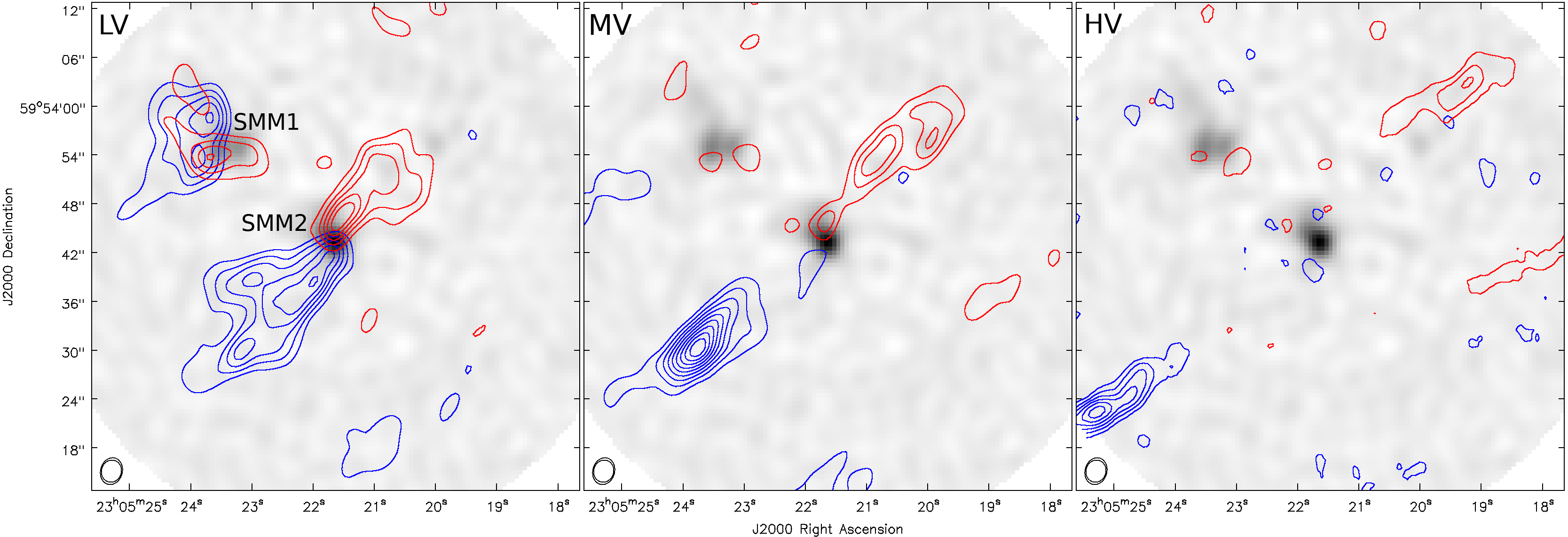}
     \caption{Integrated intensity maps of the blue- and red-shifted CO(2-1) emission (contours), overlaid on the 1.3 mm continuum emission (gray scale). 
     The intensity was integrated in three different velocity ranges: low (LV, left panel), intermediate (MV, middle panel), and high velocities (HV, right panel). We integrated from $-$55.3 to $-$67.7 \kms\ (blue) and from $-$47.3 to $-$39.3 \kms\ (red) for LV; from $-$68.4 to $-$84.4 \kms\ (blue) and from $-$38.7 to $-$27.7 \kms\ (red) for MV; and from $-$85.1 to $-$105.4 \kms\ (blue) and from $-$27 to $-$16.1 \kms\ (red) for the HV emission. In the bottom left of each panel we show the beam size of both the CO and the continuum emission. The contours represent levels starting at 10$\sigma$, and increase with 10$\sigma$ steps. 
     }
     \label{fig:CO-mom0}
 \end{figure*}

To further investigate the morphology and kinematics of the CO gas, we made integrated intensity (moment-0) maps in three different velocity ranges: low (LV), intermediate (MV), and high velocities (HV). 
In Figure~\ref{fig:CO-mom0} we show the CO(2-1) velocity-binned moment-0 maps in blue and red contours overlaid on the 1.3 mm continuum emission in gray scale\footnote{Note that in Fig.~3 of \citetalias{Birkmann07} the blue- and red-shifted lobes appear inverted compared to our results. This is due to a mislabeling of the words "dashed" and "dotted" in their caption (Linz, priv. communication).}. 
The left panel shows the LV emission, the middle panel the MV emission, and the right panel the HV emission. The detailed definitions of these velocity ranges are given in the caption of Fig.~\ref{fig:CO-mom0}. Inspection of this figure reveals a number of interesting features. First, in all velocity ranges the emission shows a clear bipolar structure, which indicates that the flow is close to the plane of the sky. However, the measured radial velocities are rather large ($\gtrsim 50\,$\kms offset from the systemic velocity), which implies some degree of inclination. These features can be explained by a rather narrow, and fast flow observed at medium inclination.
Second, the CO(2-1) emission becomes more collimated with offset velocity. In the LV panel it shows an X-shaped morphology, and in the HV panel the emission is only marginally resolved across the jet direction.
We can quantify the degree of collimation using the ratio between the major and minor axes of the emission \citep{Bally&Lada83,Wu04}. Measuring the maximum and minimum projected radii of each lobe for each velocity range we obtain collimation factors of 1.5, 3.5 and 6.5 for the LV, MV and HV range, respectively. Third, the flow velocity increases with distance from SMM2. The red and blue lobes have a projected length, measured from the source to the outermost detected emission at 3$\sigma$, of $\sim\,$0.65 and 0.8 pc, respectively, i.e. a total projected length of the CO(2-1) emission of about 1.45 pc. However, we note that the blue emission appears to extend beyond the field of view, so this number is clearly a lower limit.

\begin{deluxetable*}{l c c c c c c c c c}%[h]
\tablenum{4}
\tablecaption{Outflow parameters.\label{tab:flow-parameters}}
%\tabletypesize{\small}
\tabletypesize{\scriptsize}
\tablewidth{0pt}
\tablehead{
\colhead{}  &\colhead{$R_{proj}$} & \colhead{$M_{out}$}& \colhead{$P_{out}$} & \colhead{$E_{out}$} & \colhead{$V_{char}$} & \colhead{$t_{dyn}$} & \colhead{$\dot M_{out}$} & \colhead{$\dot P_{out}$} & \colhead{$\dot E_{out}$} \\
\colhead{} & \colhead{[pc]} & \colhead{[\msun]} & \colhead{[$10^2$ \msun\ \kms]} & \colhead{[$10^{46}$ erg]} & \colhead{[\kms]} & \colhead{[$10^4$ yr]} & \colhead{[$10^{-4}$ \msun\ yr$^{-1}$]} & \colhead{[$10^{-3}$ \msun\ \kms\ yr$^{-1}$]} & \colhead{[\lsun]}
} 
\startdata
Blue lobe & 0.8 & 25.4$\,\pm$7.0 & 2.4$\,\pm$0.7 & 2.9$\,\pm$0.8 & 9.2  & 8.2 & 3.1$\,\pm$0.9 & 2.9$\,\pm$0.8 & 2.9$\,\pm$0.8\\
Red lobe  & 0.65 & 19.7$\,\pm$5.5 & 2.0$\,\pm$0.6 & 2.3$\,\pm$0.6 & 10.3 & 6.2 & 3.1$\,\pm$0.9 & 3.2$\,\pm$0.9 & 3.2$\,\pm$0.9\\
Total & 1.45 & 45.2$\,\pm$12.6  & 4.4$\,\pm$1.2 & 5.2$\,\pm$1.4 & 9.7  & 7.2 & 6.2$\,\pm$1.7 & 6.0$\,\pm$1.7 & 5.9$\,\pm$1.6
\enddata
\end{deluxetable*}

We estimate the physical parameters of the outflow by using the CO(2-1) emission to derive the \htwo\ column density per pixel, and per channel $N_{pix}(H_2)$ using the formalism presented in \cite{Dunham14}, which assume optically thin emission and LTE conditions. 
We used for the excitation temperature a conservative value of T$_{ex}=50$ K and a CO abundance ratio of [CO]/[\htwo$]=10^{-4}$ \citep{Frerking82, Leung84}. 
Then, we estimated the corresponding mass using the expression
\begin{equation}
    M_{out}^{pix}=\mu_H m_H d^2 N_{pix}(H_2) \Omega_{pix} 
\end{equation}
where $\mu_H$ is the mean molecular weight \citep[2.8, ][]{Kauffmann08}, $m_H$ is the mass of a hydrogen atom, $d$ is the distance to the source, and $\Omega_{pix}$ is the solid angle of a pixel. Finally, the values were summed over all pixels and velocity channels. We obtained a mass of 25.5$\,(\pm7)$ and 19.7$\,(\pm5.5)$ \msun\ for the blue and red lobes, respectively. 
Following the definitions in the work of \cite{Plunkett15} we also obtained the momentum $(P_{out})$, energy $(E_{out})$, characteristic velocity $(V_{char})$, dynamical time $(t_{dyn})$, mass loss rate $(\dot{M}_{out})$, force $(\dot P_{out})$, and mechanical luminosity $(\dot E_{out})$ for each lobe and the entire flow. 
In Table~\ref{tab:flow-parameters} we show the results obtained.

\begin{figure}
    \centering
    \includegraphics[width=0.97\columnwidth]{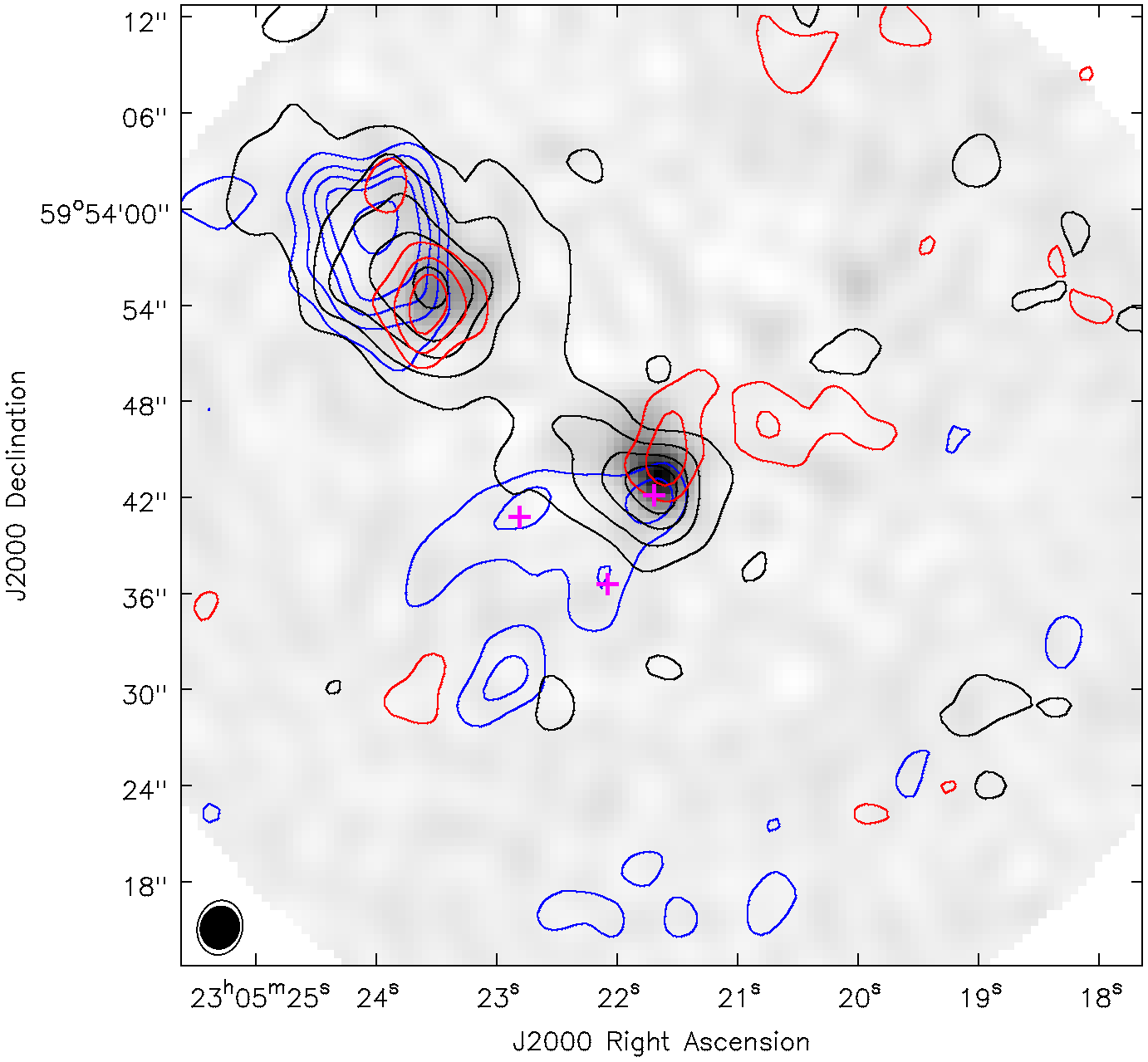}
    \caption{Integrated intensity map of the \tco(2-1) emission (contours), overlaid on the 1.3~mm continuum (gray scale). Blue and red contours show the results of integrating the \tco(2-1) emission between [$-$58.4, $-$55.4] and [$-$49.3, $-$47.0]\kms, respectively. Black contours show the emission integrated at core velocities, i.e., between $-54.6$ to $-$50.1~\kms. Contours represent 5, 10, 15, 20 and 30$\sigma$ levels (with $\sigma$ being 0.05 and 0.12~\mjy\ for the emission at wing and core velocities, respectively). The contoured and filled ellipses in the bottom left show the beam size of the molecular and continuum emission, respectively, and the magenta crosses indicate the positions where the \co/\tco\ intensity ratio was measured (see Section~\ref{subsec:discussion-uncertainties}).
    }
    \label{fig:13CO-mom0}
\end{figure}
\begin{figure}
    \includegraphics[width=0.43\textwidth]{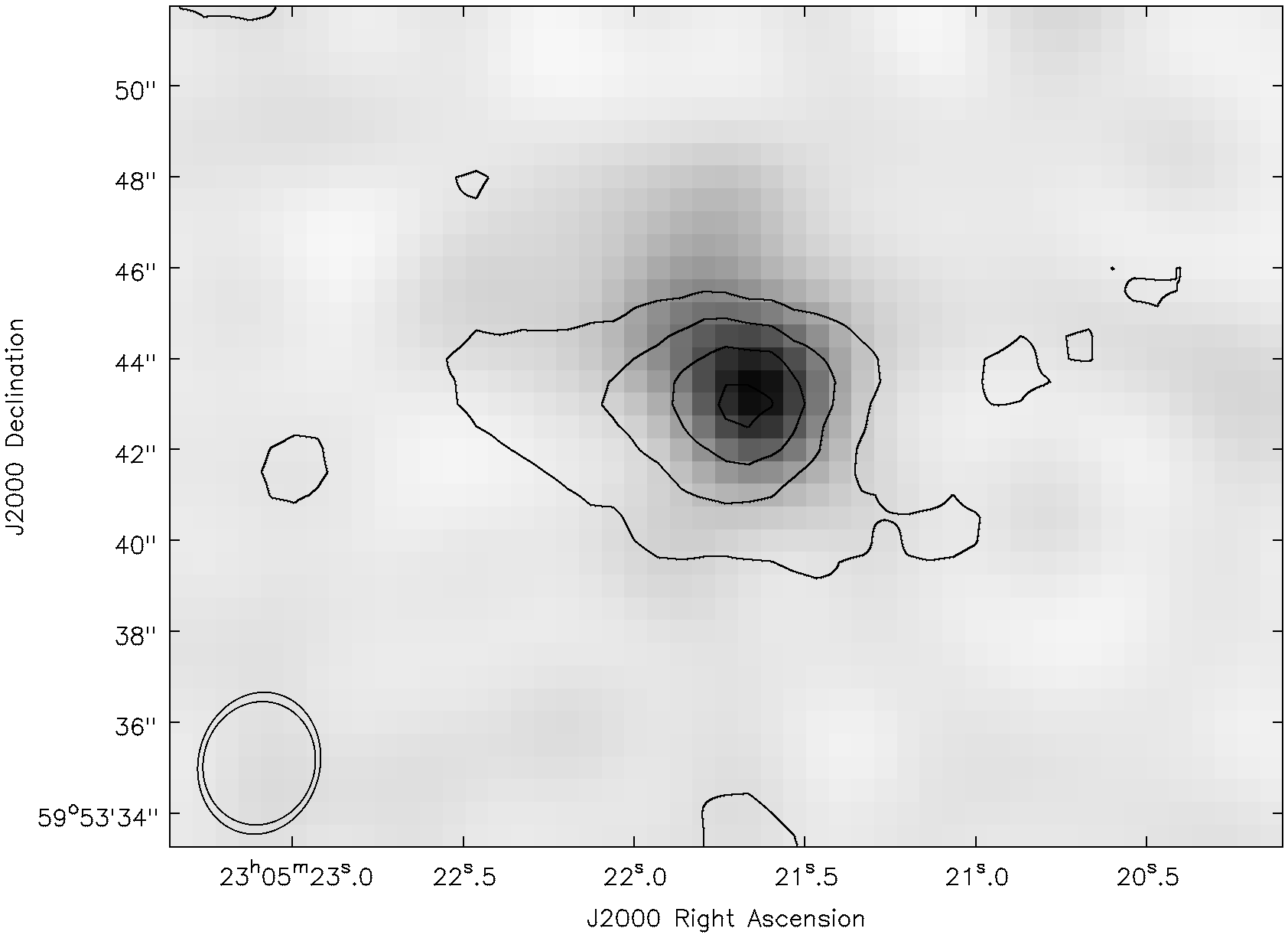}
    \includegraphics[width=0.49\textwidth]{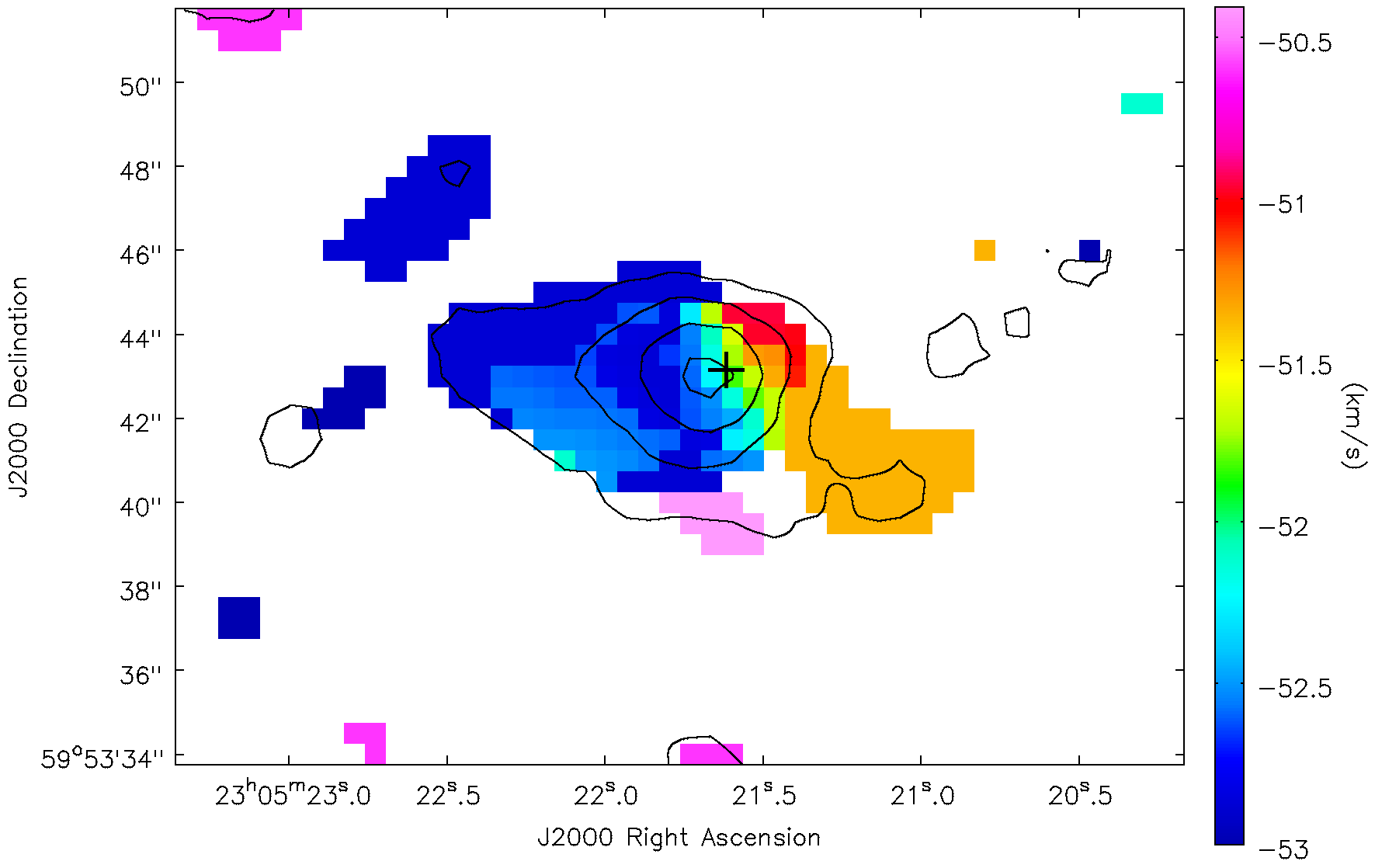}
    \caption{\textit{Top}: \ceo(2-1) integrated intensity emission (contours) toward SMM2 overlaid on the 1.3 mm dust continuum emission (gray scale). The contours show 5, 10, 15 and 20$\sigma$ levels. The ellipses in the bottom left show the 1.3 mm and the \ceo\ synthesized beam sizes. \textit{Bottom}: intensity weighted velocity field of the \ceo\ emission (color scale). The contours are the same as in the top image, and the cross marks the position of the 3.6 cm emission. We can see a \aprox~1.5~\kms\ velocity gradient in the NE-SW direction, approximately perpendicular to the outflow direction.}
    \label{fig:C18O-mom-maps}
\end{figure}

\vspace{0.1cm}

Regarding the \tco\ emission, the gas velocity range is not as wide as that of the \co, as we can see in Figure \ref{fig:CO-spectra}. We made an integrated intensity map of this isotopologue emission, which we show in Figure \ref{fig:13CO-mom0}. The blue and red contours show the distribution of the blue-shifted (between $-58.4$ and $-55.4$~\kms) and red-shifted emission (between $-49.3$ and $-47.0$~\kms), respectively, while the black contours show the result of integrating at core velocities (from $-54.6$ to $-50.1$~\kms). 
The emission in both lobes seems to be consistent with the X-shape morphology observed in the $^{12}$CO LV emission. We note as well that the \tco\ emission at core velocities traces well the structure and orientation of the filament previously studied by \citet[fig.~6]{Bihr15}, and that the outflow axis is perpendicular to the filament.

\vspace{0.1cm}

We also made an integrated intensity map of the \ceo(2-1) emission, which we show overlaid on the $1.3\,$mm continuum in Figure \ref{fig:C18O-mom-maps} (upper panel). The emission is elongated in the NE-SW direction, approximately perpendicular to the outflow direction. Fitting a 2-D Gaussian to this map we obtain a deconvolved size of $8\overset{\prime\prime}{.}4\,(\pm1^{\prime\prime})\times3\overset{\prime\prime}{.}9\,(\pm0\overset{\prime\prime}{.}7)$ ($\sim 36200 \times 16800\,$au) oriented at a PA of 80\fdeg7$\,(\pm8\overset{\circ}{.}2)$. 
The peak of the \ceo(2-1) emission is coincident with that of the $1.3\,$mm emission within our measurement accuracy. Assuming optically thin emission, LTE conditions, a temperature of $17.3$~K, and an abundance of $[\text{C}\,^{18}\text{O}]/[\text{H}_2]=1.7\times10^{-7}$ \citep{Frerking82}, the column density, density, and total mass of the C$\,^{18}$O structure are $6.2\,(\pm1.3)\times 10^{22}$~cm$^{-2}$, $1.4\,(\pm0.5)\times 10^5$~cm$^{-3}$ and $2.2\,(\pm0.8)\,$\msun, respectively.

In the bottom panel of Figure~\ref{fig:C18O-mom-maps} we show the \ceo\ intensity-weighted velocity field (moment-1) map in color scale, and overlaid the C$\,^{18}$O moment-0 contours. The cross marks the position of the 3.6 cm emission peak. We can see a \aprox 1.5~\kms\ velocity gradient in the NE-SW direction. If this velocity gradient were caused by Keplerian rotation about the center of the core, the implied central mass would be $\sim14\,$M$_\odot$.

\section{Discussion}
\label{sec:discussion}

\subsection{Uncertainties in the derived flow parameters}
\label{subsec:discussion-uncertainties}
One of the assumptions made when estimating the outflow parameters was that the \co\ emission is optically thin. However, it has been shown that, in low-\textit{J} transitions, the \co\ line-wings are generally optically thick \citep[e.g.,][]{Goldsmith84, Arce01, Curtis10b}.
%Inspecting the spectra shown in Fig.~\ref{fig:CO-spectra}, we see that when the $^{13}$CO isotopologue %is
%detected, the ratio to the $^{12}$CO emission is always less than the optically thin ratio of $62$ \citep{Langer93}, indicating significant opacity.
Hence, an opacity correction is necessary. 

It is in principle possible to estimate the \co\ optical depth $\tau_{12}$ from the relation
\begin{equation}
\frac{T_{mb,12}}{T_{mb,13}}=\frac{1-e^{-\tau_{12}}}{1-e^{\frac{-\tau_{12}}{X_{iso}}}}
\label{eq:12/13ratio}
\end{equation}
where $\frac{T_{mb,12}}{T_{mb,13}}$ is the ratio between the \co\ and \tco\ brightness temperatures, and $\tau_{12}$ is the \co\ optical depth. This equation assumes equal excitation 
temperature and filling factors for the two isotopologue transitions. 
However, the application of this method to our data is not straight forward, because we detected the $^{13}$CO line
only at low velocity offsets from the core, in a region where the main $^{12}$CO line is very bright, and residual
phase errors degrade the map quality. Furthermore, the presence of extremely strong self-absorption at low velocities
makes a detailed mapping of the $\frac{T_{mb,12}}{T_{mb,13}}$ ratio impossible. In order to get an estimate of the
optical depths involved at low velocities, we have carefully inspected the velocity cubes and selected three locations where \tco\ and \co\ are free of the problems discussed above, and have a clear velocity and position correspondence.
These positions are shown as crosses in Fig.~\ref{fig:13CO-mom0}, and we measure a nearly constant 
$\frac{T_{mb,12}}{T_{mb,13}}$ ratio of 19.2 in the aforementioned spots. Using equation~\ref{eq:12/13ratio} and an abundance ratio $X_{iso}=\frac{[^{12}\text{CO}]}{[^{13}\text{CO}]} = 62$, this translates to $\tau_{12}\approx 3$. Applying an opacity correction of the form ${\tau_{12}(1-e^{-\tau_{12}})^{-1}}$ to the mass calculation done in section \ref{subsec:results-co} results in a factor 3.2 increase, at least in the LV range. We cannot determine an opacity correction at higher velocities, hence it is unclear whether this correction factor would also apply to the MV and HV ranges.

Another source of uncertainties is the inclination \textit{i} of the flow. As mentioned above, because
the red and blue lobes do not overlap, but have rather large radial velocities, the inclination likely has an intermediate value. We can estimate an inclination correction using $i_0 =$~57\fdeg3, which corresponds to the average of an equally likely distribution of
inclinations \citep{Bontemps96}. Table~8 of \cite{Dunham14} lists the dependence with $i$ of the flow kinematic
parameters and the corresponding correction factors.
For the aforementioned average inclination $i_0$, the flow momentum $P_{out}$, energy $E_{out}$, dynamical time $t_{dyn}$, force $\dot{P}_{out}$, and mechanical luminosity $\dot{E}_{out}$ have correction factors of 1.9, 3.4, 0.6, 2.9 and 5.3, respectively. Note that the estimated mass is not affected by projection.
 
Finally, we note that we have adopted an excitation temperature of $50\,$K for the outflowing gas traced by the CO(2-1) line. This number seems reasonable despite the relatively low temperature of most of the core gas, as some shock heating in the outflow is expected. Similar numbers are also suggested by multi-line studies \citep[e.g.,][]{Qiu19}, and in the most extreme cases values up to $1000\,$K have been reported \citep{Su12}.
For our case, if the assumed excitation temperature is larger than the assumed $50\,$K that would lead to substantially larger outflow mass, and thereby result in increased values for the energetic flow parameters.

In summary, from the considerations discussed in this section, it appears that our estimates for the flow mass and energetic parameters are lower limits, and are likely higher by factors of a few.

\subsection{Flow Kinematics}

\begin{figure*}
    \centering
    \includegraphics[width=.9\textwidth]{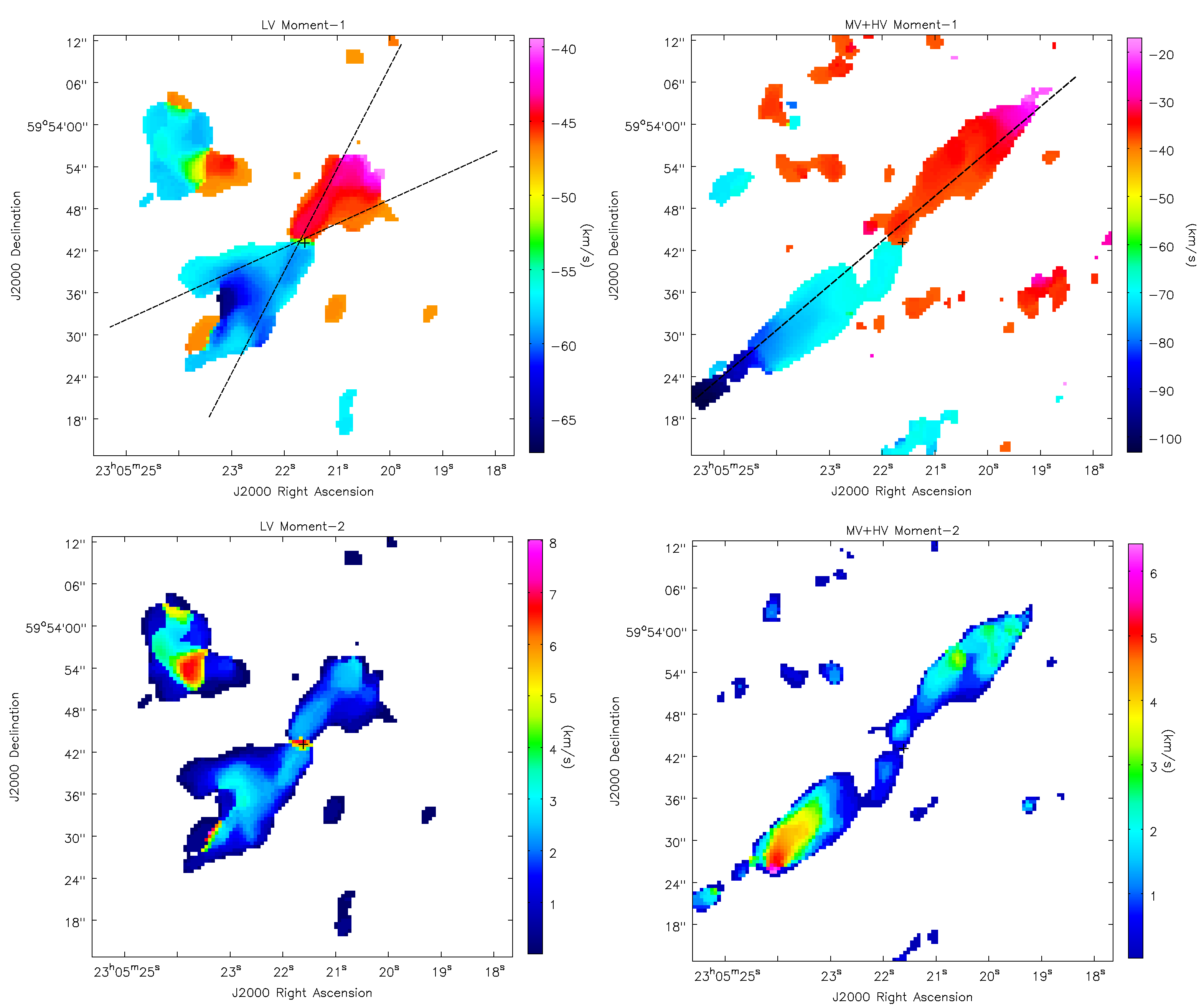}
    \caption{\textit{Top:} intensity weighted velocity field (moment-1) of the CO(2-1) line. The left panel shows the LV range, and in the right panel we show the combined MV+HV range. The dashed black lines show the directions of the PV cuts shown in Fig.~\ref{fig:CO-PV}. 
    \textit{Bottom:} intensity weighted velocity dispersion (moment-2) of the CO(2-1) line. The left panel shows the LV range, and in the right panel we show the combined MV+HV range. The black cross marks the position of the $3.6\,$cm continuum source.
    Note that for obtaining the best dynamic range, in each panel the color scale extends over a different velocity range.
    }
    \label{fig:CO-moments}
\end{figure*}

In order to further study the kinematic features of the CO flow, we made intensity weighted velocity field (moment-1) maps, integrating over the LV range, as well as over the MV and HV ranges combined (MV+HV), which we show in the top panels of Figure~\ref{fig:CO-moments}. As seen before in the channel and moment-0 maps of the CO emission, the LV emission has a conical 
appearance with an opening angle of about $40^{\circ}$, while the MV+HV emission is highly collimated, with velocity increasing with distance from the central source. We also find in the inner $10^{\prime\prime}$ a change of flow angle from $\sim$130\deg\ to about 160\deg.

In the bottom panels of Figure~\ref{fig:CO-moments} we show intensity weighted velocity dispersion (moment-2) maps of the outflow created by integrating over the same velocity ranges as before. 
In the LV emission, we can see a large velocity dispersion towards the center of SMM2 ($\sim8\,$\kms),
where with our angular resolution (corresponding to about $10000\,$au), both blue and red wings are observed. Thus, the flow launching zone is likely at this position.
Further out in the lobes, the dispersion is higher in the center of the flowing gas, and smoothly decreasing to the edges. We also note that there is a feature in the blue lobe, located at about 15\arcs\ SE from SMM2, with an abrupt increase in velocity dispersion. In the moment-1 map we can see isolated, red-shifted CO emission at this position with a velocity of about $-$48~\kms\ ($|V-V_{LSR}|\sim 5\,$\kms). While other interpretations are possible, the higher velocity dispersion could be explained by an interaction of the flow with ambient material.
Regarding the MV+HV emission, we observe a strong velocity dispersion region in the blue-shifted lobe, of about 6~\kms, that decreases gradually towards the NW.
In the red-shifted lobe we can see velocity dispersion patterns that resemble bow-shock structures. Again, these high dispersion structures could be caused by interaction of the outflow with the surrounding matter.

\begin{figure*}
    \centering
    \includegraphics[width=.32\textwidth]{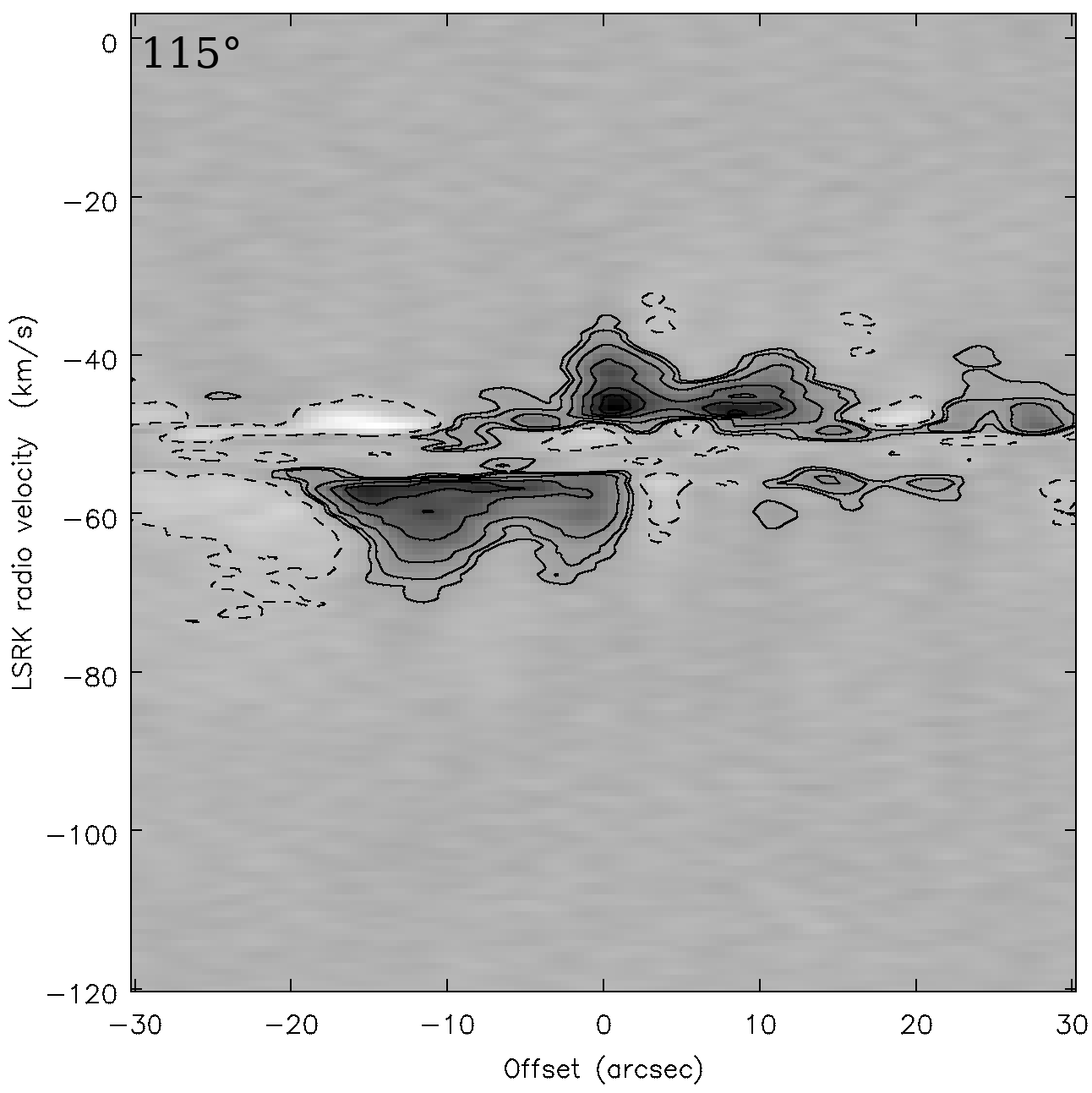}
    \includegraphics[width=.32\textwidth]{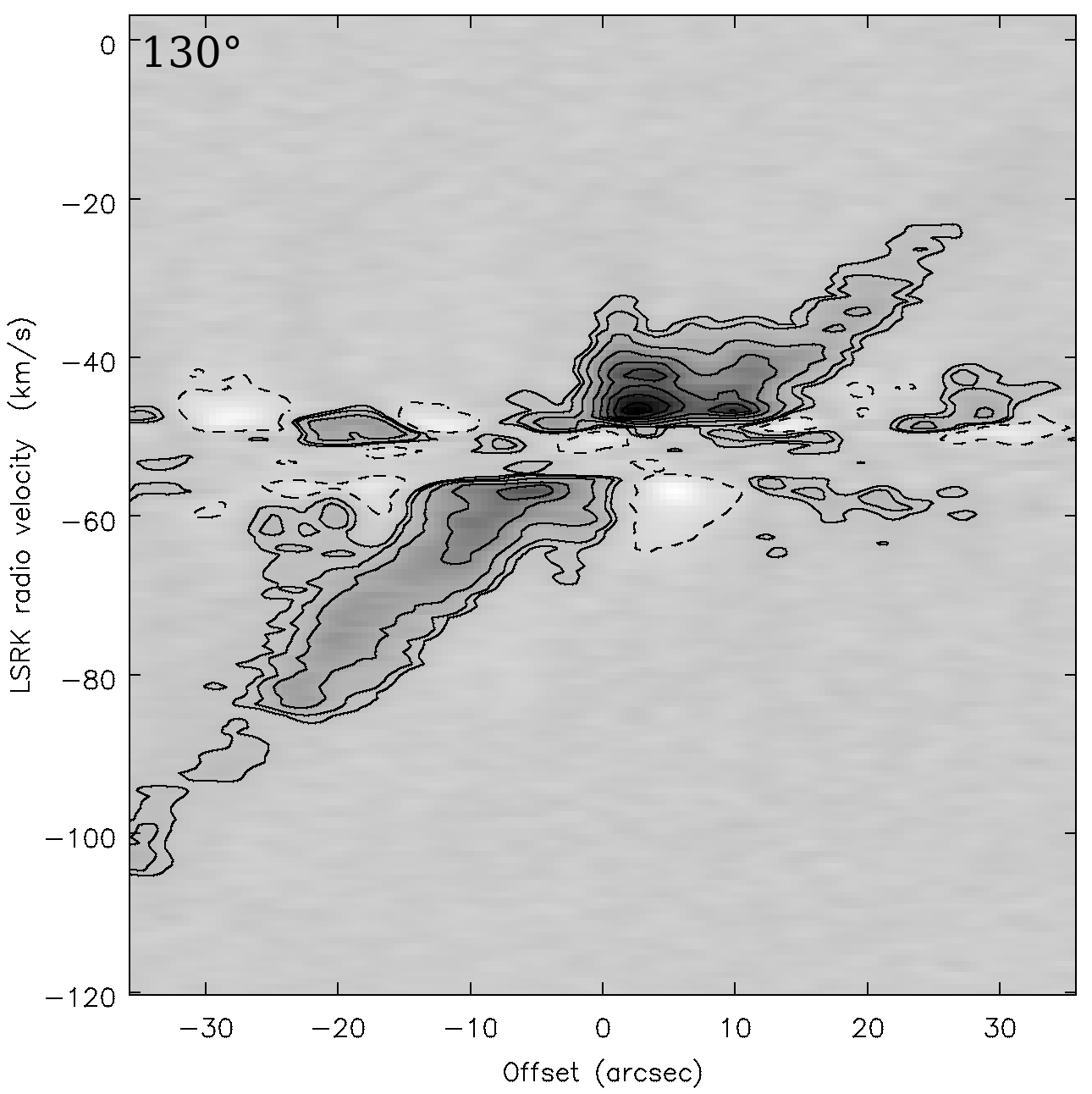}
    \includegraphics[width=.32\textwidth]{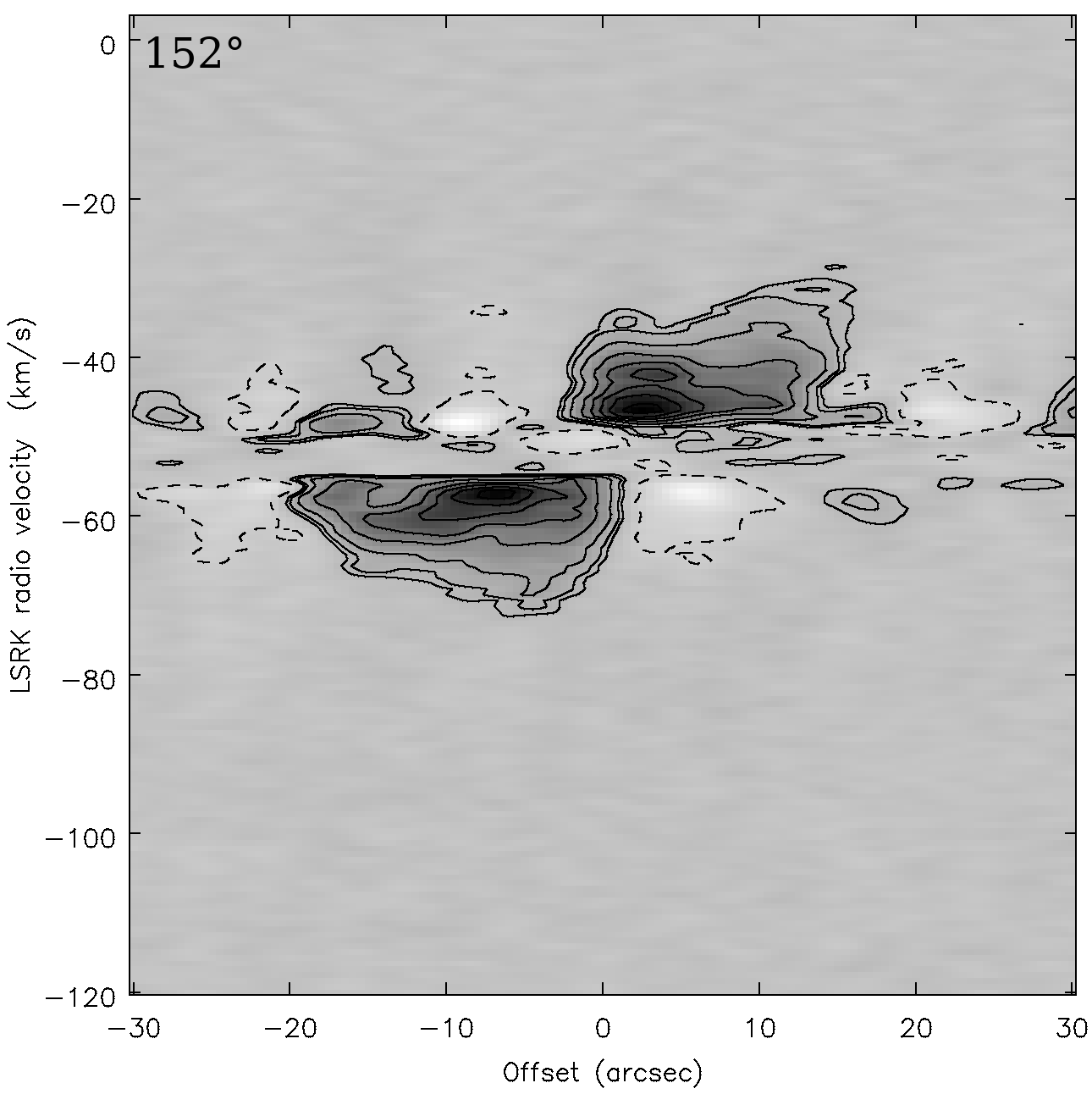}
    \caption{PV diagrams resulting from the cuts done at 115\deg\ (left), 130\deg\ (middle), and 152\deg\ (right), which we show in Fig. \ref{fig:CO-moments}.
    Note that the 115\deg\ and 152\deg\ cuts share the same offset center, while the 130\deg\ cut was done so that it would go over both the blue- and red-shifted HV $^{12}$CO emission. The contours represent $-3, 3, 5, 10\sigma$ levels and keep increasing with $10\sigma$ steps (with $\sigma=0.06$~\Jyb). We can see in the 115\deg\ and 152\deg\ PV diagrams velocity extents of 10 and 15~\kms, while the 130\deg\ cut reveals a linear increase in velocity with distance.
    }
    \label{fig:CO-PV}
\end{figure*}

To probe the velocity structure in the conical LV range, we made position-velocity (PV) diagrams along the CO flow with PAs of 115\deg and 152\deg, as well as a PV diagram along the flow with a PA of 130\deg\ to study the HV kinematic features (Fig.~\ref{fig:CO-PV}). 
The 115\deg\ cut shows spur-like structures with velocity extents of 10 to 15~\kms. Similar structures have been reported in several
flows, and appear to be consistent with a jet-driven scenario \citep[e.g.][]{Lee2001}.
The same features are also present in the 152\deg\ cut, although to a somewhat lesser extent.

The 130\deg\ PV diagram traces the main axis of the flow along the collimated jet-like component. It shows an almost linear relation between the gas position and its velocity, i.e., a Hubble flow-like behavior. The velocity gradient along this structure is about $60\,$km$\,$s$^{-1}\,$pc$^{-1}$. Such behavior has been interpreted with a jet-bow shock model with episodic ejection events (e.g. see discussion in \citealt{Arce07}). The bow shock-like features in the MV+HV moment-2 map (Fig.~\ref{fig:CO-moments}) support this interpretation. On the other hand we note that the CO emission appears relatively smooth along this axis, which suggests a more continuous ejection of matter. Another possibility to explain a smooth Hubble flow-like velocity relation was suggested for the flow in the high-mass YSO IRAS 20126+4104 \citep[see][Fig.~3]{Su2007}. In this case, the authors suggest that the linear velocity gradient is a projection effect due to a precessing flow. Since in our case
we observe a change of outflow angle (see MV+HV moment-1 map, Fig.~\ref{fig:CO-moments}), suggestive of precession, we consider this scenario a distinct possibility as well.

\subsection{The Nature of ISOSS J23053+5953 SMM2}

\citetalias{Birkmann07} have suggested that ISOSS J23053+5953 SMM2 is a promising candidate for a high-mass protostar in a very early phase of evolution. The data reported in this paper allow us to draw a number of conclusions on the nature of this object.

First, we need to address the question whether the observed CO structure, i.e., a highly collimated and fast outflow, together with a broader, lower velocity component, arises from a single driving source or whether it consists of several flows aligned on the sky due to projection. If we are observing multiple flows, we would expect to see different radial velocity components on the lobes, due to either a lack of angular resolution or an effect of superposition on the plane of the sky. An example of this phenomenon is N159W YSO-S in the LMC, where ALMA Cycle 1 observations show a bipolar flow with overlapping blue and red wings \citep[][Fig. 2]{Fukui15-flowsuperpos}, which newer Cycle 4 data resolved into several bipolar flows \citep[][Fig. 1]{tokuda19-flowsuperpos}. However, we do not see such overlapping in the CO channel maps nor moment-0 maps, and the fast/narrow and slow/wide flow components appear smoothly connected throughout the lobes. Thus, our flow data are consistent with a single central source near the position of the $3.6\,$cm continuum source.
We also note that there is evidence that such outflow structures can be produced by a single source, e.g. the class~0 YSO IRAS$\,$04166+2706, which was imaged in great detail by \cite{SantiagoGarcia09}.

Second, we can address the question of the mass of the central object considering our VLA cm data. As described in Section~\ref{subsec:VLA-results}, the cm data show a rising spectrum as one would expect from an ionized jet, and the $6\,$cm radio luminosity $S_{\nu}\,d^2$ of $1.3\,(\pm0.4)\,$mJy$\,$kpc$^2$ is typical for a high-mass object. Furthermore, the $S_{\nu}\,d^2$ vs. L$_{bol}$ data for this source falls very near the expected relation for ionized jets \citep{Anglada18, Rosero19, Purser21}. Also, based on the extensive Gould Belt survey of cm emission from low mass YSOs \citep[e.g.,][]{Dzib13}, at the sensitivity level of our observations, no low-mass YSO should be detected at the distance of ISOSS J23053+5953 SMM2 \citep[see][]{Rosero19}.
We note, though, a clear mismatch between the orientation of the CO outflow and the deconvolved PA of the cm emission (see Table 2), assumed to stem from the ionized jet. Even though this could be connected to a precessing outflow, we caution about an over-interpretation of this fact since the cm emission is weak and only mildly resolved in our data, and the PAs are similar to the orientation of the synthesized beam.

Third, a similar argument for a young high-mass star at the center of the flow can be made from considering the outflow energetic parameters. The flow mass and energetics we have measured are several orders of magnitude larger than what is normally seen toward low-mass YSOs. In particular, we estimate a lower limit of the momentum rate of $6\times10^{-3}\,$M$_\odot\,$km$\,$s$^{-1}\,$yr$^{-1}$, whereas values lower than $10^{-4}\,$M$_\odot\,$km$\,$s$^{-1}\,$yr$^{-1}$ are typical for low-mass YSOs \citep{Bontemps96}. 

Based on SED modeling, \citetalias{Birkmann07} estimated a very high value for the accretion rate of $2.1\times 10^{-3}\,$M$_\odot\,$yr$^{-1}$, which consequently leads to the expectation that a future high-mass star is forming in SMM2.
In principle, the accretion rate can be related to the mass outflow rate of a stellar wind which would provide the momentum to drive the observed CO flow, hence $P_{out} = P_w = m_w v_w $, where $P_w, m_w$ and $v_w$ are the momentum, mass, and speed of the stellar wind, respectively \citep[e.g.,][]{Zhang05}. Unfortunately we lack precise values for these latter quantities, and we only note that if we adopt $v_w = 100\,$km$\,$s$^{-1}$, $t_{dyn}\sim 1.5\times10^4\,$yr (see below), and a fraction $f$ between ejected mass by the stellar wind and accreted mass of $\approx 0.1$, our data are consistent with the accretion rate derived by \citetalias{Birkmann07}.

Thus, from the above discussion our data strongly support the suggestion of \citetalias{Birkmann07} that ISOSS J23053+5953 SMM2 is a future high-mass star in an early phase of formation.

Given this conclusion, it is interesting to investigate the age of the high-mass protostar ISOSS J23053+5953 SMM2. Below, we briefly discuss two approaches for this important question.
The first method rests on the dynamical time scale of the outflow. We obtained a value of $t_{dyn} = 7.2\times 10^4\,$yr, which is based on a characteristic velocity as described in \cite{Plunkett15}, i.e., $t_{dyn}=R_{max}/V_{char}$ with $V_{char}=P_{out}/M_{out}$.
While $V_{char}$ ($\sim$ 9~\kms) is low in comparison to the velocities the CO emission reaches, and consequently results in a relatively large time scale, a smaller value for the age of the flow results when using the largest velocity offset of the CO emission, in which case $t_{dyn}\sim 1.5\times10^4\,$yr, consistent with the results of \citetalias{Birkmann07} based on SED modeling. However, since the blue wing of the outflow extends beyond our primary beam, both values could be larger.

A second possible way of gauging the age of this system rests on the large observed difference between the core mass estimated from the \ceo\ emission (M$_{C^{18}O} \sim 2\,$M$_\odot$) and the 1.3 mm continuum emission (M$_{1.3mm} \sim 46\,$M$_\odot$). The low temperature ($\sim$17~K) and high density ($> 10^6\,$cm$^{-3}$) in the SMM2 core create a favorable environment for gas freeze-out onto dust grain surfaces \citep[e.g.,][]{Caselli99,Hernandez11,Fontani12,Bovino19}. 
Assuming optically thin emission in the \ceo\ line, our measured column densities correspond to an observed abundance of [\ceo]/[H$_2$]$\,\sim8.5\times 10^{-9}$, much lower than the "canonical" value of $1.7\times 10^{-7}$, implying a large depletion factor $f_D\sim20$. With the formation and evolution of a protostellar object at the core center, the CO depletion drops due to evaporation \citep[e.g.,][]{Giannetti14}, hence the large $f_D$ estimated hints at an extreme youth of the source. Based on our limited data, affected by uncertainties related to the \ceo\ optical depth and the dust opacity $\kappa_{\nu}$, it is difficult to obtain a quantitative value for the age of the central source, but we note that for the prevailing density in the SMM2 core a freeze-out timescale of about 5000~yr is implied \citep[see][]{Bergin07}, which is a lower limit also consistent with the model presented by \citetalias{Birkmann07}.

\section{Conclusions}
\label{sec:conclusions}

In this paper we have presented observations of the star forming core ISOSS J23053+5953 SMM2 with the VLA at 6~cm, 3.6~cm, and 7~mm, as well as with the SMA in the 1.3~mm continuum, and in the $J=2-1$ transitions of three CO isotopologues. The main results of these observations are as follows: 
\\

\noindent 1) We detect compact continuum emission in the VLA 6 and 3.6$\,$cm bands near the peak of the $1.3\,$mm SMM2 dust core. The rising spectral index ($\alpha = 0.24\,\pm0.15$),
and the $6\,$cm radio luminosity $S_{\nu}\,d^2$ of $1.3\,(\pm0.4)\,$mJy$\,$kpc$^2$ suggest that the cm emission arises from an ionized jet from a high-mass YSO.
\\

\noindent 2) The VLA $7\,$mm emission as well as the SMA $1.3\,$mm trace optically thin dust. We estimate a total mass for the SMM2 dust core of $45.8\,\pm13.4)\,$M$_\odot$.
\\

\noindent 3) The CO(2-1) main isotopologue revealed a bipolar molecular flow oriented in the SE-NW direction. The flow consists of a broad, low velocity component, plus a highly collimated, high velocity component. The flow is centered on the SMM2 dust core and is consistent with the location of the VLA cm sources.
\\

\noindent 4) We estimate a total mass of the outflowing gas of $45\,(\pm13)$M$_\odot$, and an outflow rate of $6.2\,(\pm1.7)\times 10^{-4}\,$M$_\odot\,$yr$^{-1}$, typical for flows from high-mass YSOs.
\\

\noindent 5) From the flow data we estimate the age of this system as between $1.5 - 7.2 \times 10^4\,$yr. This young age is also supported by the observed depletion of the C$\,^{18}$O(2-1) line.
\\

In summary, our observations demonstrate that ISOSS J23053+5953 SMM2 is an excellent candidate for a young high-mass object with a jet/flow system. Our study adds to the
growing evidence for the presence of both collimated jets and wide-angle winds during
the formation process of high-mass stars \citep[e.g.,][]{Torrelles11, Carrasco-Gonzalez15}. The central source is clearly in a pre-HMC phase and is likely still in the process
of assembling most of its mass.
\\
\\
\small{
\textit{Acknowledgments.} We wish to thank Dr. Riccardo Cesaroni for thoughtful discussion and comments. We also thank the anonymous referee for comments and suggestions that helped improve this work. P. H. and E. D. A. acknowledge support from NSF grants AST–1814011, and AST–1814063, respectively.
}

\bibliography{biblio}{}
\bibliographystyle{aasjournal}

\end{document}